\documentclass[conference]{IEEEtran}
\IEEEoverridecommandlockouts
\usepackage{cite}
\usepackage{amsmath,amssymb,amsfonts}
\usepackage{algorithmic}
\usepackage{graphicx}
\usepackage{textcomp}
\usepackage{xcolor}
\def\BibTeX{{\rm B\kern-.05em{\sc i\kern-.025em b}\kern-.08em
    T\kern-.1667em\lower.7ex\hbox{E}\kern-.125emX}}

\usepackage{algorithmic}
\usepackage{graphicx}
\usepackage{textcomp}
\usepackage{xcolor}

\usepackage{tabularx}
\usepackage{listings}
\usepackage{subfigure}
\usepackage{array}
\usepackage{flushend}
\usepackage[ruled,vlined]{algorithm2e}
\usepackage{framed}
\usepackage{amsmath,amsfonts,tikz,multirow,stmaryrd,xspace,enumitem}
\usepackage{url}
\usepackage{amsthm}
\newtheorem{theorem}{Theorem}
\usepackage{pgfplots}
\pgfplotsset{compat=newest}
\usepackage{ellipsis, ragged2e, marginnote}
\usepackage[]{hyperref}
\hypersetup{hidelinks}
\usepackage{threeparttable}
\usepackage{marvosym}
\usepackage{balance}
\usepackage{threeparttable}
\usepackage{diagbox}

\definecolor{mygray}{gray}{.9}
\definecolor{mypink}{rgb}{.99,.91,.95}
\definecolor{mycyan}{cmyk}{.3,0,0,0}
\definecolor{mygreen}{RGB}{1,145,1}
\definecolor{lightgreen}{RGB}{144,238,144}

\def\oursys{MapComp }
\def\oursysno{MapComp}

\def\eg{\textit{e.g.}, }
\def\ie{\textit{i.e.}, }

\newtheorem{definition}{Definition}

\def\simulator{\mathtt{Sim}}
\def\sender{\mathcal{S}}
\def\receiver{\mathcal{R}}

\def\alice{\mathcal{P}_0}
\def\bob{\mathcal{P}_1}

\def\party{\mathcal{P}}

\newcommand{\protocolsym}[1]{\mathtt{P}_{\mathtt{#1}}}
\newcommand{\functionsym}[1]{\mathtt{F}_{\mathtt{#1}}}

\newcommand{\share}[1]{\ensuremath{\langle #1\rangle}\xspace}
\newcommand{\shareA}[1]{\ensuremath{\langle #1\rangle}^{\textsf{A}}\xspace}
\newcommand{\shareB}[1]{\ensuremath{\langle #1\rangle}^{\textsf{B}}\xspace}
\newcommand{\shareb}[1]{\ensuremath{\langle #1\rangle}^{\textsf{b}}\xspace}

\newcommand{\sqlword}[1]{{\small\textsf{#1}}}
\newcommand{\tablesym}[1]{{\mathtt{#1}}}
\newcommand{\aggfun}[1]{{\mathsf{#1}}}
\newcommand{\domain}[1]{\mathbb{D}^{#1}}

\newcommand{\smalltextsf}[1]{{\small\textsf{#1}}}

\newcommand{\binaryValue}[1]{\texttt{#1}}

\newcommand{\partitle}[1]{\medskip \noindent \textbf{#1.}}

\newif\iffullversion 

\fullversiontrue

\begin{document}

\title{MapComp: A Secure View-based Collaborative Analytics Framework for Join-Group-Aggregation
}


\author{\IEEEauthorblockN{
Xinyu Peng\IEEEauthorrefmark{1}\IEEEauthorrefmark{2},
Feng Han\IEEEauthorrefmark{2}, 
Li Peng\IEEEauthorrefmark{1}\IEEEauthorrefmark{2}\textsuperscript,
Weiran Liu\IEEEauthorrefmark{2},
Zheng Yan\IEEEauthorrefmark{3},
\\ 
Kai Kang\IEEEauthorrefmark{2}, 
Xinyuan Zhang\IEEEauthorrefmark{2}, 
Guoxing Wei\IEEEauthorrefmark{2},  
Jianling Sun\IEEEauthorrefmark{1},
Jinfei Liu\IEEEauthorrefmark{1},
Lin Qu\IEEEauthorrefmark{2}
}
\IEEEauthorblockA{
\IEEEauthorrefmark{1}Zhejiang University, 
\IEEEauthorrefmark{2}Alibaba Group, 
\IEEEauthorrefmark{3}Xidian University
\\
\IEEEauthorrefmark{1}\{steven.pengxy, jerry.pl\}@alibaba-inc.com, \{sunjl, jinfeiliu\}@zju.edu.cn, 
\\
\IEEEauthorrefmark{2}\{fengdi.hf, weiran.lwr, pufan.kk, xinyuan.zxy, guoxing.wgx\}@alibaba-inc.com, xide.ql@taobao.com,
\\
\IEEEauthorrefmark{3}zyan@xidian.edu.cn
}}

\maketitle

\begin{abstract}

Join-group-aggregation (JGA) queries are fundamental to data analytics, yet executing them collaboratively across different parties poses significant privacy risks. Secure multi-party computation (MPC) offers a cryptographic solution. However, existing MPC-based JGA approaches consider only a one-time query paradigm and suffer from significant performance bottlenecks. It executes expensive join operations from scratch across multiple queries and employs inefficient group-aggregation (GA) protocols, both of which hinder their practical use for scalable, real-time analysis.

This paper introduces MapComp, a novel view-based framework to facilitate JGA queries for secure collaborative analytics. Through specially crafted materialized views for join and novel design of GA protocols, MapComp removes duplicate join workload and expedites subsequent GA, improving the efficiency of JGA query execution. To address the challenge of continuous data updates, our materialized view offers \textit{payload-independence} feature and provides significant efficiency improvements in view refreshing with \textit{free} MPC overhead. This feature, on the other hand, also allows further acceleration for GA, where we devise multiple novel protocols that outperform prior works. 
Notably, our work represents the \textit{first} endeavor to expedite secure collaborative JGA queries using materialized views. 
Our rigorous experiments demonstrate a significant advantage of MapComp, achieving up to a $308.9\times$ improvement in efficiency over the baseline in real-world query simulations. Moreover, our optimized GA protocols achieve up to a $1140.5\times$ improvement compared to prior oblivious sorting-based solutions.
\end{abstract}


\vspace{-0.1in}
\section{Introduction}



Data analysis has become indispensable for numerous businesses seeking valuable insights. Since massive valuable data is collected 
and held within different parties (\eg Amazon, Google), there is significant potential for data holders to derive great benefits by collectively analyzing their private datasets \cite{IonKNPSS0SY20,bater2017smcql}. 
Among statistical data analysis queries, join-group-aggregation (JGA) queries are particularly crucial for extensive applications including market investigation \cite{fent2023practical}, advertisement conversion analysis \cite{lepoint2021private}, 
online auction analysis \cite{tran2007transformation}, and constitutes a substantial portion of real-world analytics \cite{JohnsonNS18}. The JGA query paradigm involves first performing a join over input data, followed by a sequence of group-aggregation (GA) processes. However, conducting a collaborative JGA query may pose significant privacy threats.

\vspace{-0.05in}
\partitle{Motivating Example} An online advertisement supplier (AS) tracks users who have clicked on a particular advertisement for a product, and has a data table $\tablesym{ad}(\mathsf{userId}, \mathsf{productId}, \mathsf{clickDate})$.
A product company (PC) places advertisements on AS and knows which users have made purchases, and maintains a data table $\tablesym{order}(\mathsf{userId}, \mathsf{productId}, \mathsf{orderAmount})$. AS may want to compute the total sales brought by the advertisement each day, i.e., the total amount spent by users who made a corresponding purchase after seeing an advertisement:

\vspace{0.03in}
\framebox{\begin{minipage}{0.9\linewidth}
\sqlword{\textbf{SELECT} $\tablesym{ad}$.$\mathsf{clickDate}$, sum($\tablesym{order}$.$\mathsf{orderAmount}$) \\
\textbf{FROM} $\tablesym{ad}$ \textbf{JOIN} $\tablesym{order}$
\textbf{ON} $\tablesym{ad}$.$\mathsf{userId}$ = $\tablesym{order}$.$\mathsf{userId}$ \\
\textbf{AND} $\tablesym{ad}$.$\mathsf{productId}$ = $\tablesym{order}$.$\mathsf{productId}$ 
\textbf{GROUP BY} $\tablesym{ad}$.$\mathsf{clickDate}$
}
\end{minipage}}
\vspace{0.03in}



The challenge in evaluating this JGA query is that the data are held separately by two parties as their private data, whose sharing is restricted by privacy concerns and regulations \cite{GDPR,HIPAA}. To address this challenge, a promising paradigm is to perform secure queries with privacy-preserving techniques such as secure multi-party computation (MPC) to ensure end-to-end privacy \cite{bater2017smcql}.

This paper tackles two critical performance bottlenecks that hinder the practical application of MPC-based collaborative JGA queries, especially in multi-query scenarios. First, the join workload in JGA queries are both \textit{highly repetitive} and \textit{computationally expensive}.  Analytical workloads often involve multiple JGA queries operating on the same joined table with various ad-hoc GA processes, making the join process duplicated. However, current MPC-based solutions follow the one-time processing paradigm \cite{poddar2021senate,wang2021secure,liagouris2021secrecy,han2022scape}, executing each join independently and preventing the reuse or pre-computation of costly workloads. Second, the subsequent GA is inherently inefficient in multiple queries. Recent works \cite{poddar2021senate,wang2021secure,liagouris2021secrecy,han2022scape} rely on the oblivious sorting-based approach, which incurs substantial overhead due to its $O(nl \log n)$ \cite{hamada2012practically} complexity, where $n$ is the size of data and $l$ is the bit-length of values. These two performance bottlenecks severely limit the efficiency and applications of secure JGA queries. \textit{To this end, can we design a JGA query framework that reduces the duplicated join overhead and enables a faster GA process?}

\vspace{-0.0in}
\partitle{Challenges} 
To address the above efficiency problems, adopting a pre-computed \textit{materialized view} \cite{SrivastavaDJL96} to cache and reuse intermediate join results and a new design of GA protocols hold promise. 
Nevertheless, designing a materialized view for secure join processes poses several challenges. First, a secure materialized view must be refreshed for updates in the payload (the attributes in addition to the join keys, such as group-by and aggregation attributes) before it is reused. In practical scenarios, databases undergo dynamic and frequent updates due to business growth or status changes (\eg client account balances are continually updated).
However, previous MPC-based join solutions fail in lightweight view refreshing. They either requires re-execution from scratch upon any payload update \cite{wang2021secure,rindal2021vole}, or consider payload dynamics with heavy view refresh operations \cite{zhou2024shortcut}.
Second, the subsequent GA processing over the materialized join view requires careful design to ensure compatibility with the view's data structure and optimization for efficiency. 
However, existing works that consider payload dynamics either suffer from heavy view refresh operations \cite{zhou2024shortcut} or only support limited subsequent query types \cite{WangBNM22} (\ie join-select queries). 

\vspace{-0.0in}
\partitle{Contributions}
In this paper, we introduce \oursysno, a novel framework that supports a view-based query processing paradigm to expedite secure JGA queries. Inspired by the idea of using a join index as a materialized view \cite{valduriez1987join}, \oursys designs a specially crafted materialized view to eliminate duplicated join workload and expedite subsequent GA queries. To address the challenge of efficiently refreshing the view in response to payload dynamics, our designed view decouples the join alignment process from the data payload (\ie using only the join keys for alignment). This feature we call \textit{payload-independence}. 
Consequently, while prior works that rely on a payload-dependent process with at least $O(n)$ MPC operations \cite{wang2021secure,zhou2024shortcut,rindal2021vole}, our view allows \textit{MPC-free} view refreshing (\ie without requiring MPC operations) without compromising security, making it exceptionally efficient and practical for dynamic databases.
To construct our materialized view, we introduce a novel secure primitive called Alignment-PSI (private set intersection). Alignment-PSI obliviously generates permutations that align intersection elements by carefully exploiting MPC building blocks, including circuit PSI and shuffling, which may be of independent interest. 

In addition to eliminating redundant joins, our proposed view improves the performance of subsequent GA queries. 
With the help of payload-independence, the payload inputs for GA remain in plaintext. This allows us to further optimize GA protocols by partial local processing and leveraging insights into the inherent characteristics of plain data. We improve conventional oblivious sorting-based approaches by using stable sorting and/or bitmaps to perform group dividing, resulting in a more efficient GA for multiple cases of group cardinality. 

Experimental results show that \oursys outperforms existing solutions without a view-based paradigm, achieving up to a $308.9\times$ improvement in efficiency over the baseline in a real-world query simulation. Moreover, our optimized GA protocols achieve up to a $1140.5\times$ improvement compared to prior oblivious sorting-based solutions.

%




\vspace{-0in}
\section{MapComp Overview}
\label{sec:overview}
\vspace{-0.05in}


\vspace{-0in}
\subsection{System Overview}
\vspace{-0.05in}

\begin{figure}[t!]
    \centering
    \includegraphics[width=0.75\linewidth]{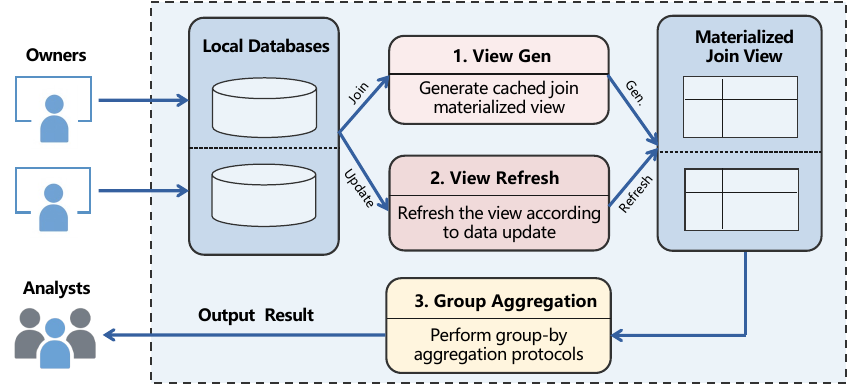}
    \vspace{-0.1in}
    \caption{System architecture and workflow.}
    \vspace{-0.2in}
    \label{fig:system-workflow}
\end{figure}
The system architecture and workflow of \oursys are shown in Fig. \ref{fig:system-workflow}. 
\oursys involves two roles: (1) two mutually distrustful data owners $\alice, \bob$, and (2) the analyst. Data owners have private data stored in local databases, and will coordinately execute a secure protocol to compute the result of a query issued by the analyst. The analyst is agreed upon by all parties before any computation starts, and a data owner can also play the role of an analyst. In the rest of our paper, we treat $\bob$ as the analyst to describe our protocols for simplicity.
Similar to \cite{wang2021secure}, we assume the database schema, the query, and the size of input databases are public knowledge. 



\vspace{-0in}
\subsubsection{Workflow}

Initially, the data owners identify the data to be involved in JGA queries and run the view generation process to generate materialized join views. The views are cached to accelerate follow-up JGA queries. Subsequently, when the original data is updated, the data owner can initiate a view refresh to keep the view up-to-date, thereby ensuring the GA protocol on it computes the correct query results. 
Upon receiving the analyst's online ad-hoc JGA query, data owners can run the GA protocols directly on the corresponding materialized join view to obtain the aggregation result. Since the views have already been generated, the duplicated join is eliminated, and the whole workload is sped up. 

It is worth mentioning that the view generation workload is independent of subsequent GA in the JGA query, so it can be performed at any preferred time before JGA running, \eg offline, to minimize online overhead. 
The refresh strategies can be determined according to requirements, \eg refresh only after receiving a query request.



\vspace{-0in}
\subsubsection{Query formulation}
This paper focuses on the JGA queries, which cover a considerable number of real-world statistical queries \cite{JohnsonNS18,fent2023practical,IonKNPSS0SY20,lepoint2021private,tran2007transformation}. The pattern of JGA is described below.


\vspace{0.04in}
\framebox{\begin{minipage}{0.9\linewidth}
\sqlword{\textbf{Select} $\tablesym{R^0}.g_0$, $\tablesym{R^1}.g_1$, $\aggfun{agg}_0$($\tablesym{R^0}.v_0$), $\aggfun{agg}_1$($\tablesym{R^1}.v_1$)}\\
\sqlword{\textbf{From} $\tablesym{R^0}$ \textbf{Join} $\tablesym{R^1}$ \textbf{on} $\tablesym{R^0}.k = \tablesym{R^1}.k$ \textbf{Group by} $\tablesym{R^0}.g_0, \tablesym{R^1}.g_1$;}
\end{minipage}}
\vspace{0.05in}

Consider two data tables $\tablesym{R^0}, \tablesym{R^1}$ owned by $\alice, \bob$ respectively. We abbreviate the above JGA query as $\mathcal{G}_{(g_0, g_1), \aggfun{agg}(v_0, v_1)}(\tablesym{R^0} \Join_k \tablesym{R^1})$,
where $\tablesym{R^0} \Join_k \tablesym{R^1}$ means $\tablesym{R^0}$ join $\tablesym{R^1}$ on $\tablesym{R^0}.k = \tablesym{R^1}.k$ and $k$ is the join attribute. 
The above query first partitions the join result according to grouping attributes $(g_0, g_1)$, and then computes the aggregate functions $\aggfun{agg}(\cdot)$ over the aggregation attributes $v_0, v_1$ for each group, where $\aggfun{agg}(v_0, v_1)$ is the abbreviation of $\{\aggfun{agg}_u(v_u)\}_{u\in\{0, 1\}}$, and $\aggfun{agg}_u$ can be $\aggfun{max}$, $\aggfun{min}$, $\aggfun{sum}$, and $\aggfun{count}$. 
Multiple join or group attributes from one party can be concatenated together as input and we denote them as a single attribute for simplicity. The expression evaluation of multiple attributes before GA, \eg $\aggfun{agg}(v_0 \times v_1)$, is naturally covered since it can be realized with plain/encrypted pairwise computation. 

\subsection{System Features}

\partitle{Materialized-view based query acceleration} 
Mapcomp employs a specially crafted materialized view to efficiently support JGA across multiple queries. Our materialized view can be reused and precomputed to accelerate subsequent workloads, even when the data payload is continuously updated. Compared with baselines \cite{wang2021secure,zhou2024shortcut,rindal2021vole} that require at least $O(n)$ complexity for view refreshing for join, our view enjoys free (zero) MPC overhead. Thus, the existence of materialization allows duplicate join workloads to be completely avoided and, for the first time, facilitates MPC analysis over real-time updating databases. For GA workload, MapComp exploits the payload-independence feature of views again and improves prior GA solutions by up to $1140.5\times$ (see \S\ref{sec:evaluation} for details).

\partitle{Broad coverage of common SQL operators}
MapComp provides extensive support for common SQL operators, with a primary focus on accelerating JGA queries essential for statistical analysis. It accommodates the most critical join types, including PK-PK, PK-FK, and multi-column equi-joins, which collectively represent the vast majority of joins found in real-world workloads \cite{JohnsonNS18}. Similarly, its support for aggregation functions, such as COUNT, SUM, MAX, MIN, and AVG, covers over 99\% of practical use cases \cite{JohnsonNS18}. Other relational operators (\eg ORDER BY) can be directly supported by sequential combinations of secure sub-protocols (\eg a secure ORDER-BY protocol \cite{liagouris2021secrecy}) or local processing. 


\partitle{Privacy preservation} Mapcomp aims to support collaborative querying on the joint database of both parties without revealing any plain or intermediate sensitive data to any party. This is ensured by the security guarantee of the underlying MPC primitives (see \S\ref{sec:sec_model} below). MapComp defends against semi-honest adversaries and can be easily extended to malicious security settings (see \S\ref{sec:discussion}). 



\vspace{-0in}
\subsection{Security Model}
\label{sec:sec_model}
\vspace{-0in}

We focus on the standard semi-honest two-party computation (2PC) model, which is a commonly adopted model in a majority of prior MPC-based query systems \cite{WangBNM22,bater2017smcql,wang2021secure,BaterP0WR20,bater2018shrinkwrap}.
We consider a static probabilistic polynomial-time (PPT) semi-honest adversary, who can corrupt at most one computing party.
The adversary may attempt to learn information from the transcript (\ie all messages sent and received during the protocol) of the protocol while following the prescribed protocol.
The security of our protocols is defined under the simulation paradigm. We refer to \cite{lindell2017simulate} for the formal definition of security and omit it here.

Our protocols are a sequential composition of existing building blocks whose individual security has been proven, and all intermediate results between those protocols are in secret-shared form (see \S\ref{sec:secretShare}). 
Thus, the security of our protocols can be proven with the sequential composition theorem \cite{lindell2017simulate}.
We omit the detailed proof, as this is a standard framework in the security literature \cite{demmler2015aby,wang2021secure}.

\vspace{-0in}
\section{Preliminaries}
\label{sec:preliminaries}
\vspace{-0.05in}
\vspace{-0.05in}
\subsection{Notations}
\vspace{-0.05in}

We denote $[a,b]$ as $\{a, a+1\ldots,b\}$ for $a \leq b$ and $[a]$ is shorthand for $[1,a]$.
$ind(\phi)$ is an indicator function which outputs 1 when $\phi$ is $\binaryValue{true}$ or 0 otherwise.
We use $\perp$ to represent $\binaryValue{null}$ and assume $\aggfun{agg}(\perp, x) = x$ holds for any aggregate function $\aggfun{agg}$ and input $x$.
We denote $X = (x_1, \dots, x_n)$ as a vector with length $|X| = n$, and specifically let $x_j = \perp$ for $j > |X|$ to simplify the description in the follow-up. 
We denote $X||Y$ as the vector $(x_1 || y_1, \dots x_n || y_n)$.
We denote a permutation $\pi: [n] \rightarrow [n]$ as $\pi = (\pi(1), \pi(2), \dots, \pi(n))$. 
$Y = \pi \cdot X$ means applying $\pi$ on a vector $X$, which outputs $Y = (x_{\pi(1)}, \ldots, x_{\pi(n)})$. 
For example, applying $\pi = (2,3,1)$ on $X = (a, b)$ will output $Y = (b, \perp, a)$. 


Given a table $\tablesym{T}$, we denote $\tablesym{T_i}$ as its $i^{\text{th}}$ tuple, $\tablesym{T}[v]$ as the vector containing all values of the attribute $v$, and $\tablesym{T_i}[v]$ as the $i^{\text{th}}$ value of $\tablesym{T}[v]$. 
We denote $\domain{g}$ as the domain space of attribute $g$, and $\domain{g}_i$ as the $i^{\text{th}}$ distinct value in $\domain{g}$. 
We denote the group cardinality $d_0 = |\domain{g_0}|, d_1 = |\domain{g_1}|$, respectively. We use the letter with a subscript or superscript $u \in \{0,1\}$ to represent a value held by $\party_u$. 

A bitmap that encodes attribute $g$ consists of bit-vectors of number $|\domain{g}|$, each of which orderly represents whether tuples equal to $\domain{g}_j$ for $j \in [|\domain{g}|]$. For example, an attribute $g$ with $|\domain{g}| = 2$ and tuples $(\mathsf{red}, \mathsf{blue}, \mathsf{red}, \mathsf{blue})$ can be encoded as 2 bit-vectors $(b^{\mathsf{r}}, b^{\mathsf{b}})$, where $b^{\mathsf{r}}=(1,0,1,0)$, $b^{\mathsf{b}}=(0,1,0,1)$.

Other frequently used notations are shown in Appendix \ref{app:notations}.

\vspace{-0.1in}
\subsection{Secret Sharing}
\label{sec:secretShare}
\vspace{-0.0in}




A 2-out-of-2 additive secret sharing scheme splits a secret value $x$ into $x_0$ and $x_1$ with the constraint that $x = x_0 + x_1 \bmod 2^k$ for $x \in \mathbb{Z}_{2^k}$ (i.e. $\shareA{x}_i = x_i$ is an arithmetic share) or $x = x_0 \oplus x_1$ for $x \in \mathbb{Z}_2^k$ (i.e. $\shareB{x}_i = x_i$ is a binary share). 
Evaluating an arithmetic multiplication gate ($\shareA{x} \cdot \shareA{y}$) or a binary AND gate ($\shareB{x} \odot \shareB{y}$) requires precomputed multiplication triples \cite{beaver1992efficient}.

%



We denote $\shareb{x}$ as the binary share of a bit $x\in\{0, 1\}$, and $\shareb{\lnot x}$ as $\shareb{1 - x}$. We denote $\functionsym{B2A}, \functionsym{b2A}, \functionsym{A2B}$ as functionalities to convert between a binary share and an arithmetic share. Since the type conversions can be efficiently realized \cite{demmler2015aby,knott2021crypten}, we simply write $\share{x} = (\share{x}_0, \share{x}_1)$, ignoring the sharing type for a uniform description unless the sharing type is specified. 
We additionally require the following primitives, whose realizations are described in \cite{demmler2015aby,rathee2020cryptflow2,mohassel2018aby3}:

\vspace{-0in}
\begin{itemize}
    \item $\functionsym{mul}(f, x)$ takes a bit $f$ and $x$ from two parties, respectively, and returns $f \,?\, \share{x} : \share{0}$.
    \item $\functionsym{mux}(\shareb{f}, \share{x}, \share{y})$ returns $\share{r}$ where $r = f \,?\, x:y$. 
    \item $\functionsym{eq}(\share{x}, \share{y})$ returns $\shareb{ind(x = y)}$.
\end{itemize}

\vspace{-0in}
\subsection{Required Secure Primitives}
\label{sec:obli_primitives}
\vspace{-0in}

\noindent \textbf{Oblivious switch network (OSN).}
Assume the receiver holds a size-$n$ permutation $\pi$, and a length-$n$ vector $X$ is held by the sender or is secret-shared between two parties, OSN will output a secret-shared length-$n$ vector $\share{Y}$ which satisfies $y_i = x_{\pi(i)}$ for $i\in[n]$.
For simplicity, we denote them as $\share{Y}\leftarrow\functionsym{osn}^\textsf{s}(\pi, X)$ \cite{mohassel2013hide} and $\share{Y}\leftarrow\functionsym{osn}^\textsf{s}(\pi, \share{X})$ \cite{wang2021secure}.
They can be instantiated by Bene{\v{s}} network \cite{benevs1964optimal} with $O(n\log n)$ communication and $O(1)$ rounds.

\noindent \textbf{Random shuffle.}
$\functionsym{shuffle}(\share{X})$ randomly samples a permutation $\pi$ and permute $\share{X}$ into $\share{Y} = \share{\pi \cdot X}$.
It can be realized by invoking $\functionsym{osn}^s$ twice \cite{chase2020secret}.


\noindent \textbf{Oblivious permutation.} Suppose two parties hold a shared permutation $\share{\pi}$ and a shared vector $\share{X}$, $\functionsym{perm}^\mathsf{s}$ \cite{chida2019efficient} 
obliviously permutes $\share X$ based on $\share \pi$ and outputs $\share{\pi\cdot X}$.
The reverse version of $\functionsym{perm}^\mathsf{s}$ is $\functionsym{invp}^\mathsf{s}$ which outputs
$\share{\pi^{-1} \cdot X}$. 

\noindent \textbf{Oblivious stable sorting.} A sorting algorithm is \textit{stable} if two items with the same keys appear in the same order in the sorted result as they appear in the input \cite{asharov2022efficient}. A stable sorting $\functionsym{sSort}(\share{X})$ takes \share{X} as input and outputs $(\share{\pi}, \share{Y})$, where $Y$ is the stable sorting result of $X$, and $y_i = x_{\pi(i)}$. 
We use oblivious quick sorting \cite{hamada2012practically,wang2024relational,luo2023secure} in our implementation, which takes $O(\log l\log n)$ rounds and $O(nl\log n)$ bits of communication to sort $n$ elements with $l$-bit length.


\noindent \textbf{Permutation for one-bit vector.}
Given a secret-shared one-bit vector $\shareb{V}$, $\functionsym{perGen}(\shareb{V})$ generates a shared permutation $\share{\pi}$ representing a stable sort of $V$. For example, the permutation representing a stable sort of $(\share{1}^1, \share{0}^2, \share{1}^3, \share{0}^4)$ is $(\share{3}, \share{1}, \share{4}, \share{2})$, and applying $\pi^{-1}$ on $V$ can obtain its (obliviously) stable sorting result $(\share{0}^2, \share{0}^4, \share{1}^1, \share{1}^3)$.
Its underlying protocol \cite{chida2019efficient,asharov2022efficient} takes $O(1)$ round and the communication cost of $O(n\log n)$ bits.


\noindent \textbf{Oblivious traversal.} 
$\functionsym{trav}$ \cite{han2022scape} takes two length-$n$ shared vectors $\share{X}, \share{V}$ and an aggregate function $\aggfun{agg}$ as input, traverses and aggregates $\share{V}$ based on $\share{X}$ in a oblivious way, and outputs a length-$n$ vector $\share{Y}$ which satisfies $y_i = \aggfun{agg}(\{v_j\}_{j = \mathsf{first}(X, i)}^i)$ for $i\in[n]$. $\aggfun{agg}$ is a function whose basic unit satisfies the associative law. For example, when $\aggfun{agg} = \aggfun{sum}$, the basic unit is addition and satisfies $(a+b)+c=a+(b+c)$.
$\mathsf{first}(X, i)$ is the index of the first element within the group specified by $X_i$
. For example, given $X=(\underline{b,b},\underline{a,a,a},\underline{b,b})$, $\mathsf{first}(X, 4) = 3$, $\mathsf{first}(X, 7) = 6$.

\noindent \textbf{Circuit PSI with payload (CPSI).}
$\functionsym{cPSI}$ takes $X$ from receiver $\receiver$ and $(Y, \tilde{Y})$ from sender as input, where $|X| = n_x, |Y| = |\tilde{Y}| = n_y$ and $\tilde{Y}$ is denoted as the payload associated with $Y$.  
Informally, $\functionsym{cPSI}$ outputs two length-$n_x$ vectors $\shareb{E}$ and $\share{Z}$ to parties, where $e_i = 1$, $z_i$ = $\tilde{y_j}$ if $\exists y_j \in Y$, s.t. $y_j = x_i$, and $e_i = 0$, $z_i$ = $\perp$ otherwise.
$\functionsym{cPSI}$ hides the intersection and the payload, allowing further secure computation without revealing additional intermediate information. We refer to \cite{rindal2021vole} for formal definition and instantiation.

\vspace{-0in}
\section{Secure Materialized View}
\label{sec:view}
\vspace{-0.05in}
We present the definition of our secure materialized view and illustrate our design to generate and refresh it.

\subsection{Overview of Secure Materialized View}

The design of our materialized view is rooted in a key insight: a join operation fundamentally consists of matching and aligning tuples based on their join keys. Inspired by \cite{valduriez1987join}, we observe this alignment process can be durably captured by a join index, which materializes the mappings between input tables and the joined result. Building on this, our approach securely generates these mappings (permutations $\pi$) using only the join keys. This achieves a critical property we term payload-independence. Unlike prior works that encrypt the entire tuple (both join keys and payload) \cite{wang2021secure,zhou2024shortcut,rindal2021vole}, our method decouples the payload from the alignment logic, allowing it to remain in plaintext. The resulting benefits are profound. Any payload update triggers an extremely efficient, MPC-free view refresh, as it only requires local modifications without costly cryptographic protocols. Simultaneously, because the payload inputs to the GA workload are now in plaintext, we can devise novel and highly optimized GA protocols, as detailed in \S\ref{sec:group}. Consequently, our payload-independent view accelerates the entire JGA workflow, from join pre-computation to final GA.

\vspace{-0.05in}
\subsection{Structure of View for PK-PK join}
\label{sec:revisit}
\vspace{-0.05in}



We propose the data structure of our (PK-PK join) materialized view and defer the details of the supporting for PK-FK join in \S\ref{sec:pk-fk}.

\begin{figure}[t!]
    \centering
    \includegraphics[width=0.9\linewidth]{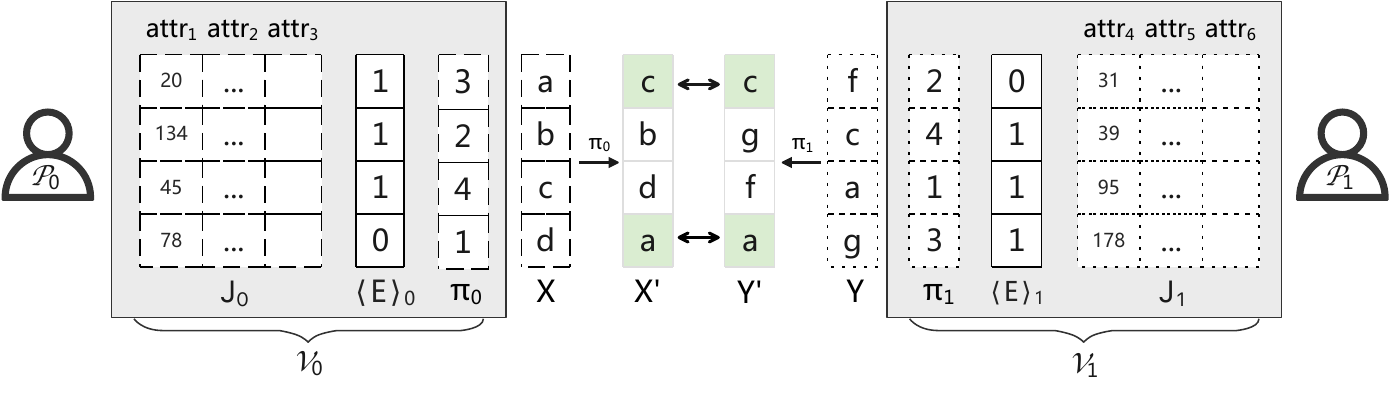}
    \vspace{-0.05in}
    \caption{A toy example of our materialized view, where $E = \share{E}_0 \oplus \share{E}_1 = (1, 0, 0, 1)$ is the intersection flag that indicates the first and fourth elements in $\textsf{X'}$ and $\textsf{Y'}$ (highlighted in green) are in the intersection. The vectors $\textsf{X'}$ and $\textsf{Y'}$ are derived by applying permutation $\pi_0$ and $\pi_1$ to the original join keys $\textsf{X}$ and $\textsf{Y}$, respectively. The flag vector $E$ is secret-shared between the two parties as $\share{E}_0$ and $\share{E}_1$. }
    \vspace{-0.2in}
    \label{fig:mapExample}
\end{figure}

\vspace{-0in}
\begin{definition}
\label{def:pkpk_view}
Given tables $\tablesym{R}^0$ with join column values $X$ and $\tablesym{R}^1$ with join column values $Y$ owned by two parties $\alice$ and $\bob$ respectively, a materialized (join) view $\mathcal{V}_u $ held by $\mathcal{P}_u$ ($u \in \{0, 1\}$) is defined as:

\vspace{-0.1in}
\[ \mathcal{V}_u =(\pi_u,\share{E}_u^b, \tablesym{J}^u) \]
\vspace{-0.15in}

\noindent where $|E| = n_e = max(|X|, |Y|)$. 
$\pi_0: [n_e] \rightarrow [n_e]$ and $\pi_1: [n_e] \rightarrow [n_e]$ are two permutations that map an index from $[n_e]$ to an index from the original data.
$\share{E}_u^b$ is the binary secret shares of intersection flag vector $E$. They satisfies $|\{e | e\in E, e = 1\}| = |X \cap Y|$ and $x_{\pi_0(i)} = y_{\pi_1(i)}$ iff $e_i = 1$.
$\tablesym{J}^u  = \pi_u \cdot \tablesym{R}^u $,  which means $\tablesym{J}^u$ is a re-ordered data transcript of $\tablesym{R}^u$ such that tuple $t_i \in \tablesym{J}^u$ equals to $t'_{\pi[i]} \in \tablesym{R}^u$ for $i \in [|X|]$.
\end{definition}



The idea behind this design is straightforward. For each equivalent pair $(x, y)$ in $X \cap Y$ ($x\in X, y\in Y$ and $x=y$), applying mapping $\pi_0, \pi_1$ on $X, Y$ respectively maps $x, y$ into the same position $i$ with $e_i = 1$. It is easy to see that $\tablesym{J}^0$ and $\tablesym{J}^1$ are aligned since they have already been permuted based on $\pi_0,\pi_1$ respectively, and the PK-PK join is essentially achieved.  For subsequent GA $\mathcal{G}_{g, \aggfun{agg}(v)}$, $\tablesym{J}^u $ can be fed directly into secure protocols to get the desired output of $\mathcal{G}$. Therefore, as long as $\mathcal{V}_u$ is consistent with the latest $\tablesym{R}^u$, the correctness of the query output under $\tablesym{J}^u$ is ensured.  In addition, the intersection flag $E$ indicates whether each tuple has been joined. It is in secret-shared form to prevent information leakage while enabling later secure computation. An example is shown in Fig.\ref{fig:mapExample}. 

\vspace{-0in}
\subsection{View Generation}
\label{sec:viewGen}
\vspace{-0.05in}
In this section, we first present a new secure primitive - Alignment-PSI and then explain how to construct our view $\mathcal{V}$ based on it.



The ideal functionality of Alignment-PSI, denoted as $\functionsym{aPSI}$ and detailed in Fig. \ref{function:aPSI}, produces an intersection flag $E$ accessible to both parties, along with a permutation $\pi_1$ provided exclusively to the receiver $\bob$. The permutation $\pi_1$ satisfies the condition $x_i = y_{\pi_1(i)}$ for all $x_i \in X \bigcap Y$. Consequently, when $\bob$ applies $\pi_1$ to his dataset, the elements within the intersection are effectively aligned. In contrast to the circuit PSI with payload $\functionsym{cPSI}$, $\functionsym{aPSI}$ introduces the additional output of $\pi_1$ while omitting the direct handling and alignment of payloads. This design allows the alignment of payloads to be performed \textit{locally} by applying $\pi_1$ to the \textit{plaintext} payload. It decouples the computation of the alignment $\pi_1$ from the subsequent alignment of payloads into entirely independent sub-steps. As a result, a single invocation of $\functionsym{aPSI}$ suffices to accommodate \textit{any} updates to the payload without incurring additional MPC-related overhead. This opens up the possibility for constructing a join view that is highly efficient in terms of payload updating. Note that $\functionsym{aPSI}$ requires $n_x \geq n_y$, and the case of $n_x < n_y$ can be covered by reversing the roles of sender and receiver.


\begin{figure}\small
    \framebox{\begin{minipage}{0.97\linewidth}
        \begin{trivlist}
            \item \textbf{Input:} The sender $\alice$'s set $X = (x_1, \ldots, x_{n_x})$. The receiver $\bob$'s set $Y = (y_1, \ldots, y_{n_y})$. $n_x \geq n_y$.
            \item \textbf{Functionality:} Upon receiving $X$ from $\alice$, $Y$ from $\bob$.

            \vspace{-0in}
            
            \begin{enumerate}
                \item Uniformly samples $E^0, E^1 \in \{0, 1\}^{n_x}$ such that $e^0_i \oplus e^1_i  = 1$ if $x_i \in X\bigcap Y$, and $e^0_i \oplus e^1_i = 0$ otherwise.
                \item Uniformly samples a size-$n_x$ permutation $\pi_1$ such that $x_i = y_{\pi_1(i)}$ for all $x_i \in X\bigcap Y$.
                \item Outputs $E^0$ to $\alice$ and $E^1$, $\pi_1$ to $\bob$.
            \end{enumerate}
        \end{trivlist}
    \end{minipage}}
    \vspace{-0.1in}
    \caption{Functionality for Alignment-PSI $\functionsym{aPSI}$.}
    \vspace{-0.1in}
    \label{function:aPSI}	
\end{figure}


\partitle{High-level idea} We describe the high-level design of our approach to implement $\functionsym{aPSI}$. Consider $\pi_1$ output by $\functionsym{aPSI}$ meets two conditions: it suffices $x_i = y_{\pi_1(i)}$ for all $x_i \in X \bigcap Y$ and it forms a permutation (\ie all elements are filled with unique integers from $[n]$). Based on these requirements, the generation of $\pi_1$ can be decomposed into two sub-tasks.


The first sub-task is to identify the positions of the intersection elements between $X$ and $Y$ and assign their corresponding indices. It can be achieved by invoking $\functionsym{cPSI}$, where $\bob$ acts as the sender and provides a payload consisting of the index vector $(1, 2, \ldots, n)$. The second sub-task focuses on assigning random values to the remaining positions while ensuring that $\pi_1$ remains a valid permutation. Our solution is to map the indices of non-intersection elements in $Y$ to the positions of non-intersection elements in $X$ by leveraging permutation operations. This ensures that the resulting vector $\pi_1$ satisfies both the intersection condition and the permutation property.

\begin{figure}[t!]
\small
    \framebox{\begin{minipage}{0.97\linewidth}
        \begin{trivlist}

            \item \textbf{Input:} The sender $\alice$'s set $X = (x_1, \ldots, x_{n_x})$. The receiver $\bob$'s set $Y = (y_1, \ldots, y_{n_y})$. $n_x \geq n_y$.
            \item \textbf{Protocol:} 
            
            \vspace{-0in}
            \begin{enumerate}
                \item Invoke $\functionsym{cPSI}$. $\alice$ acts as receiver with input $X$ and $\bob$ acts as sender with input $(Y, O)$, where $O=(1, 2, \dots, n_y)$. Then, the parties get $\shareb{E}, \share{Z}$.
                
                \item Invoke $\functionsym{cPSI}$, where $\alice$ acts as sender with input $(X, \emptyset)$ and $\bob$ acts as receiver with input $Y$. the parties get $\shareb{F'}$.

                \item If $n_x > n_y$, the parties extend $\share{F'}$ into a length-$n_x$ shared vector by padding shared zeros at the end of it.

                \item Invoke $\functionsym{shuffle}$. $(\shareb{F}, \share{L}) \leftarrow \functionsym{shuffle}(\shareb{F'}, \share{O'})$ where $O' = (1, 2, \dots, n_x)$.

                \item Compute $\share{\sigma_1}\leftarrow\functionsym{perGen}(\shareb{F})$, $\share{P^1} \leftarrow \functionsym{invp}^\mathsf{s}(\share{\sigma_1}, \share{L})$.
                
                \item Compute $\share{\sigma_0}\leftarrow\functionsym{perGen}(\shareb{E})$,  $\share{P^0}\leftarrow\functionsym{perm}^\mathsf{s}(\share{\sigma_0}, \share{P^1})$.
                \item Compute a shared permutation $\share{\pi_1}$ with $\share{\pi_1(i)} = \functionsym{mux}(\shareb{e_i}, \share{z'_i}, \share{p^0_i})$ for $i\in[n_x]$, and reveal $\pi_1$ to $\bob$.
            \end{enumerate}
            \vspace{-0in}
            
            \item \textbf{Output:} $\alice$ returns $\shareb{E}_0$ and $\bob$ returns $\pi_1, \shareb{E}_1$.
        \end{trivlist}
    \end{minipage}}
    \vspace{-0.05in}
    \caption{Alignment-PSI protocol $\protocolsym{aPSI}$.}
    \vspace{-0.05in}
    \label{protocol:mapping}	
\end{figure}

\begin{figure}[t!]
\centering
\includegraphics[width=0.9\linewidth]{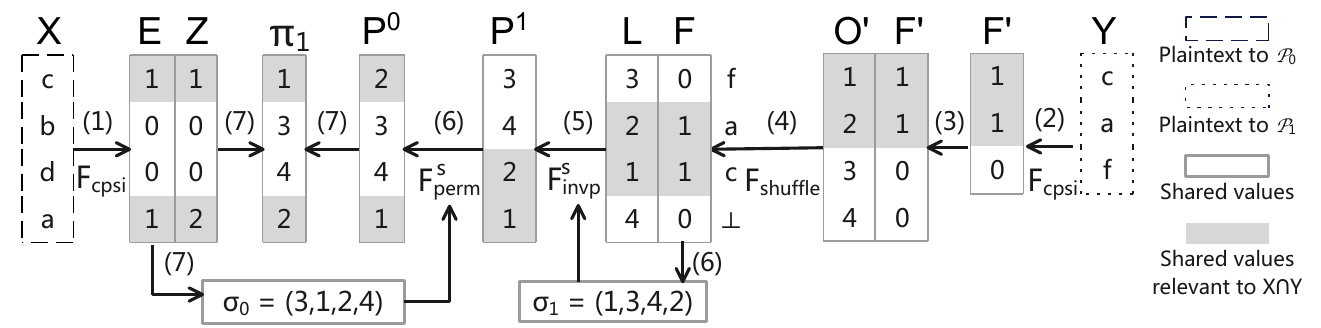}
\vspace{-0.05in}
\caption{A running example of $\protocolsym{aPSI}$ where $X = (c, b, d, a), Y = (c, a, f), n_x = 4, n_y = 3$.}
\vspace{-0.2in}
\label{fig:mapping-example}
\end{figure}

\partitle{Concrete design} We describe our protocol $\protocolsym{aPSI}$ in Fig. \ref{protocol:mapping}. The first sub-task is fulfilled by the step 1. As a result, for $i \in[n_x]$, if $e_i = 1$, $z_i$ is the index of $y_j$ that equals to $x_i$, and $z_i = 0$ otherwise. 
The second sub-task is fulfilled by the remaining part (steps 2-7).
Specifically, to identify the non-intersection elements in $Y$, the parties first perform circuit PSI in step 2 to obtain $F'$. After that, if $n_x > n_y$, the parties pad zeros into $F'$ so that the length of $F'$ equals $n_x$.
Next, to guarantee the randomness of $\pi_1$, the parties shuffle the index vector $O'$ in step 4. 
If we track a value $d\in[n_x]$ that $d > n_y$ or $y_d\notin X \bigcap Y$, $d$ will be shuffled to a random position $i$ in $L$ where $F_i = 0$ after step 4.
Then, through the correlation between $E$ and $F$, the parties map each value $l_i$ with $f_i = 0$ to a position $j$ satisfying $e_j = 0$ to obtain $P^0$.
To do that, it is observed that the numbers of 1 in $E$ and $F$ are the same, so the parties can first sort $L$ in the order of $F$ to obtain $P^1$, and then treat $P^1$ as the result of sorting $P^0$ in the order of $E$.
In particular, the parties compute a shared permutation $\sigma_1$ representing a stable sorting of a one-bit vector $F$, and applying $\sigma_1^{-1}$ on $L$ will sort $L$ in order of $F$. 
As shown in Fig. \ref{fig:mapping-example}, $P^1 = (3^{\underline{0}}, 4^{\underline{0}}, 2^{\underline{1}}, 1^{\underline{1}})$.
To reverse the sorting of $P^1$ in the order of $E$, the parties compute a shared permutation $\sigma_0$ representing a stable sorting of $E$ and apply $\sigma_0$ on $P^1$ to obtain $P^0$. 
In this way, the required permutation mentioned in the solution to the second sub-task can be seen as $\sigma_0 \cdot \sigma_1^{-1}$.
Finally, $\pi_1$ is obtained by invoking \smalltextsf{MUX} gates in step 7.

\iffullversion
We state the security of the above protocol in Theorem \ref{theorem:map}.
\else 
We state the security of the above protocol in Theorem \ref{theorem:map} and defer the brief proof to Appendix \ref{proof:map}.
\fi

\vspace{-0.05in}
\begin{theorem}
    $\protocolsym{aPSI}$ securely realizes the ideal functionality $\functionsym{aPSI}$ against a semi-honest adversary in the $(\functionsym{cPSI}, \functionsym{shuffle}, \functionsym{perGen}, \functionsym{invp}^s, \functionsym{perm}^s, \functionsym{mux})$-hybrid world.
    \label{theorem:map}
\end{theorem}
\vspace{-0.05in}

\iffullversion

\begin{proof} 
We show that there exists a PPT simulator $\simulator_0$ that can generate the simulated view given $\alice$’s inputs, which is statistically indistinguishable from the joint distribution of corrupted $\alice$’s view in the real execution of the protocol.
$\simulator_0$ can sample random strings $\shareb{\tilde{E}}_0, \share{\tilde{Z}}_0, \shareb{\tilde{F}'}_0, \shareb{\tilde{F}}_0, \share{\tilde{L}}_0, \share{\tilde{\sigma_0}}_0, \share{\tilde{\sigma_1}}_0, \share{\tilde{P}^0}_0, \\ \share{\tilde{P}^1}_0, \share{\tilde{\pi_1}}_0$ as the simulated view. 
By $\simulator_0$ running the simulator of the invoked functionalities in sequence, we can prove that the distributions of the simulated view and the real view of $\alice$ during the execution of the protocol are indistinguishable.
Similarly, the simulator $\simulator_1$ for $\bob$ can also be constructed and the security of $\protocolsym{aPSI}$ is proven.
\end{proof}

\fi


\partitle{From $\functionsym{aPSI}$ to view generation}
Note that the permutation $\pi_1$ generated by $\functionsym{aPSI}$ aligns $\bob$'s payload with $\alice$'s data. However, this alignment may introduce potential privacy risks if $\alice$'s data is structured according to a specific order (\eg sorted). To mitigate such risks, it is necessary to ensure that the intersection data of both parties is aligned to a common location that is independent of either party's data. Consequently, both parties must hold a permutation $\pi_i$, as formally defined in Definition \ref{def:pkpk_view}, to achieve this alignment while preserving privacy.

Our view generation proceeds as follows: (1) $\alice$ generates a random permutation $\pi_0$ and permutes his join key $X$ with it. This disrupts the distribution of $X$ and prevents potential privacy leakage. (2) The parties invoke $\functionsym{aPSI}$. (3) The parties permute their database with $\pi_0, \pi_1$ respectively and append the result table to the view, and finally obtain the view $\mathcal{V}$. Note that the payload of both parties is not required to be input into $\functionsym{aPSI}$ to obtain $\pi_i$, and remains plaintext (payload-independence). This feature allows efficient view refresh and accelerated subsequent GA workloads, and we will demonstrate later.

\vspace{-0in}
\subsection{View Refresh} 
\label{sec:viewRefresh}
\vspace{-0in}

With the help of the \textit{payload-independence} feature of $\mathcal{V}$, an update of the original data payload simply requires accessing and updating $\tablesym{J}$ based on the latest $\tablesym{R}$ accordingly to refresh $\mathcal{V}$. The view refresh $\protocolsym{VR}$ proceeds as follows: Upon with a payload update set $\tablesym{R}^{\text{new}}=\{i_j,t_{j}^{\text{new}}\}_{j \in [n_{\text{new}}]}$ that contains the index and content of updated tuples, for each $j \in [n_{\text{new}}]$, access the $\pi(i_j)$ -th tuples of $\tablesym{J}$ and update it with $t_{j}^{\text{new}}$.

\partitle{Efficiency} $\protocolsym{VR}$ only requires memory IO to access and modify existing tuples. It does not involve any MPC operation, making the refresh extremely efficient. We will therotically analyze and confirm 

\partitle{Security} $\protocolsym{VR}$ does not necessitate any communication with the other party. Thus, any intermediate information, \eg data update pattern \cite{WangBNM22}, is directly protected.

\vspace{-0in}
\subsection{Supporting for PK-FK Join}
\label{sec:pk-fk}
\vspace{-0in}

\iffullversion

To support PK-FK join, we build the PF-FK join view upon the PK-PK join view with additional steps. Recall that the values of an FK join key are not unique, so they can be divided into groups. Our high-level idea is to first align a single tuple within an FK group with the corresponding value of the PK. Then, we can obliviously duplicate the payloads of PK tuples to the correct locations to align the remaining tuples, so PK-FK join is achieved. The single-tuple alignment process can be reused, so the view refreshing is partially free. 

Now, we describe our design for the PK-FK join view. The definition of it differs slightly from that of the PK-PK view, and its generation and refresh require additional steps. 

W.L.O.G, we assume $\bob$'s join key is a foreign key, which means the values of $\tablesym{R^1}[k]$ are non-unique, and we can divide them into groups based on distinct FK values. Our high-level idea is to first align a single tuple within an FK group with the corresponding tuple having the same PK key. Then, the payloads of PK tuples are obliviously duplicated to the correct locations to align the remaining tuples, completing the PK-FK join. The single-tuple alignment process is independent of the payload, which means it is reusable when the payload is updated, so the view refreshing is partially free. We illustrate the definition and operations of the PK-FK join view as follows.

\vspace{-0in}
\subsubsection{View for PK-FK join}

Given two tables $\tablesym{R}^0, \tablesym{R}^1$ with join key $k$, the views held by $\party_0, \party_1$ are 
$\mathcal{V}_0 =(\pi_0, \share{E}_0^b, \share{\tablesym{J}^0}_0)$ and $\mathcal{V}_1 =(\pi_1, \sigma, \share{E}_1^b, \share{\tablesym{J}^0}_1, \tablesym{J}^1)$:

\vspace{-0.05in}
\begin{enumerate}
    \item $\tablesym{J}^1 = \sigma \cdot \pi_1 \cdot \tablesym{R}^1$, and $e_i = 1$ iff $\tablesym{J}^1_i[k] \in \tablesym{R}^0[k]$.
    \item For $1 \leq i  < |\tablesym{J}^1|$, $\tablesym{J}^1_i[k] \leq \tablesym{J}^1_{i + 1}[k]$.
    \item For $i \in [|\tablesym{J}^1|]$: if $e_i = 1$, let $p = \mathsf{first}(\tablesym{J}^1[k], i)$, then $\tablesym{J}^0_p = \tablesym{R}^0_{\sigma \cdot \pi_0(p)}$ and $\tablesym{J}^0_i = \tablesym{J}^0_p$; if $e_i = 0$, $\tablesym{J}^0_i = \tablesym{R}^0_{\sigma \cdot \pi_0(i)}$.
\end{enumerate}

\vspace{-0in}
\subsubsection{Generation and Refresh}
\label{sec:pkfkGenRefresh}

The view generation proceeds as follows, and the view refresh only requires the last two steps, so it is partially free. Note that the cost of refresh is relevantly small since it only requires oblivious switching and oblivious traversal taking $O(1)$ rounds and $O(nl \log n)$ bits of communication.

\noindent \textbf{1. Mapping and alignment.}
First, we align a single tuple within an FK group with the corresponding tuple having the same PK key. To achieve this, a constant 1 is appended to the PK value by $\alice$ and a counter number $\mathbf{t}[s]$ (\eg 1,2,3...) is appended to the FK value by $\bob$ for each tuple $\mathbf{t}$, such that $\mathbf{t}[s]$ denotes an incremental counter of tuples with the same join key value $\mathbf{t}[k]$ (within the same FK group). 
Then, the parties invoke PK-PK view generation protocols (described in \S\ref{sec:viewGen}) with inputs $\{\mathbf{t}[k] || 1\}_{\mathbf{t}\in \tablesym{R^0}}$ and $\{\mathbf{t}[k] || \mathbf{t}[s]\}_{\mathbf{t}\in \tablesym{R^1}}$, respectively. $\alice$ obtain $\pi_0$, $\shareb{E}_0$ and $\bob$ obtain $\pi_1$, $\shareb{E}_1$. 
Finally, two parties reorder the databases with $\pi_0, \pi_1$ to obtain a temporary transcript $\tablesym{D}^i = \pi_i \cdot \tablesym{R}^i$.
The tuple $\mathbf{t}^1 \in \tablesym{D}^1$ with $\mathbf{t}^1[s] = 1$ will be aligned with a tuple $\mathbf{t}^0 \in \tablesym{D}^0$ with $\mathbf{t}^0[k] = \mathbf{t}^1[k]$ if $\mathbf{t}^1[k] \in \tablesym{D}^0[k]$; or a tuple $\mathbf{t}^0 \in \tablesym{D}^0$ with $\mathbf{t}^0[k] \notin \tablesym{D}^1[k]$ otherwise. At this point, the first tuple of each FK group of $\tablesym{D}^1$ is correctly joined with the corresponding PK tuple of $\tablesym{D}^0$.

\noindent \textbf{2. Local sorting and oblivious switch.}
$\bob$ sorts the table $\tablesym{D}^1$ based on the key attributes $k, s$ to get the permutation $\sigma$ and result table $\tablesym{J}^1$. The parties invoke $\functionsym{osn}^p$ to switch $\tablesym{D}^0, \shareb{E}$ with $\sigma$ and obtain $\share{\tablesym{J}^0}, \shareb{E'}$. After this step, the tuples of $\tablesym{J}^1$ with the same key will be mapped together and sorted by $s$. 

\noindent \textbf{3. Duplicate the tuples.} To achieve PK-FK alignment, the last step is to obliviously set the payload of remaining tuples of $\share{\tablesym{J}^0}$ as correct values. The parties obliviously duplicate the tuples of $\share{\tablesym{\tablesym{J}^0}}$, such that $\tablesym{J}^0_i = \tablesym{J}^0_{\mathsf{first}(\tablesym{J}^1[k], i)}$ holds if $e'_i = 1$, where $\mathsf{first} (\cdot, i)$ returns the first index of the group $i$.

\begin{enumerate}
    \item For $i \in |\tablesym{J}^0|$, $\share{\tablesym{J}^0_i} \leftarrow \functionsym{mux}(\shareb{e'_i}, \share{\tablesym{J}^0_i}, \share{\perp})$;
    \item $\shareb{E} \leftarrow \functionsym{trav}(\share{\tablesym{J}^1[k]}, \shareb{E'}, \aggfun{xor})$, $\share{\tablesym{J}^0} \leftarrow \functionsym{trav}(\share{\tablesym{J}^1[k]}, \share{\tablesym{J}^0}, \aggfun{sum})$.
\end{enumerate}

$\alice$ set $\mathcal{V}_0 = (\pi_0, \shareb{E'}_0, \share{\tablesym{J}^0}_0)$, $\bob$ set $\mathcal{V}_1 = (\pi_1, \shareb{E'}_1, \share{\tablesym{J}^0}_1, \tablesym{J}^1)$.
This is the desired PK-FK join view output, since for every valid tuple $\tablesym{J}^1_i$ that satisfies $e_i=1$, the payload of tuple $\tablesym{J}^0_i$ is correctly joined and aligned with it.

\begin{figure}[t!]
    \centering
    \includegraphics[width=0.8\linewidth]{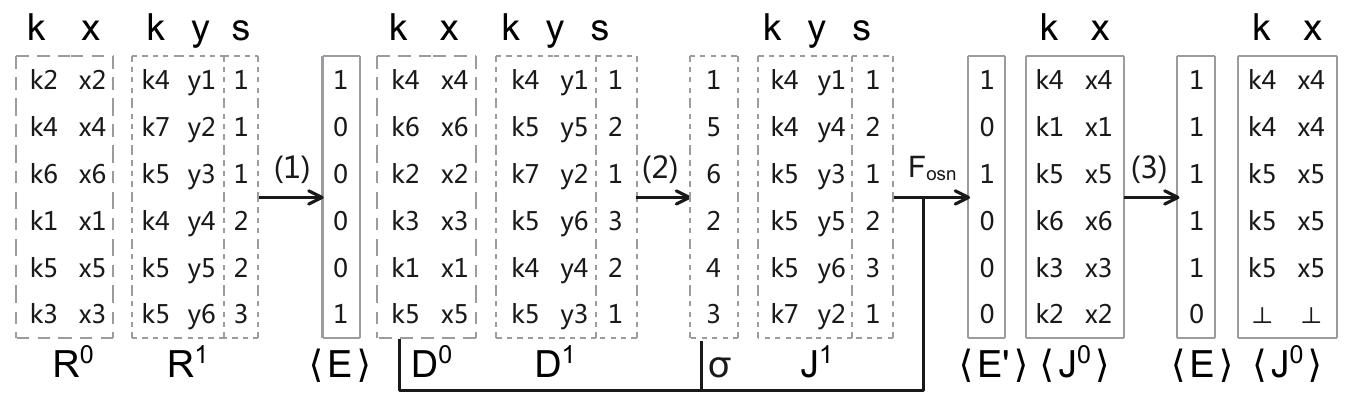}
    \vspace{-0.05in}
    \caption{Example of the view generation for the PK-FK join, where the result permutations in the first step are $\pi_0 = (2, 3, 1, 6, 4, 5)$ and $\pi_1 = (1, 5, 2, 6, 4, 3)$. We use $x, y$ to denote the other attributes besides $k$ in $\tablesym{R}^0$ and $\tablesym{R}^1$.}
    \label{fig:pkfk_view_exp}
    \vspace{-0.05in}
\end{figure}

\else

We build our PK-FK join view upon the PK-PK join view with additional steps. Recall that the values of an FK join key are not unique, so they can be divided into groups. Our high-level idea is to first align a single tuple within an FK group with the corresponding value of the PK. Then, we can obliviously duplicate the payloads of PK tuples to the correct locations to align the remaining tuples, so PK-FK join is achieved. The single-tuple alignment process can be reused, so the view refreshing is partially free. Due to space limitations, please refer to Appendix \ref{sec:app_pk_fk_join} for details.

\fi

\vspace{-0in}
\subsection{Complexity analysis}
\label{sec:viewComplexity}
\vspace{-0in}

\begin{table}[t]
    \renewcommand{\arraystretch}{1.3}
    \centering
        \resizebox{1\columnwidth}{!}{
    \begin{tabular}{|c|| c|c|| c|c|}
        \hline
        \multirow{2}{*}{\textbf{Protocols}} & \multicolumn{2}{c||} { \textbf{PK-PK join} } & \multicolumn{2}{c|} { \textbf{PK-FK join}} \\
        \cline{2-3}\cline{4-5}
        ~ & \textbf{View Gen.} & \textbf{View Refresh} & \textbf{View Gen.} & \textbf{View Refresh} \\
        \hline
        \textbf{Baseline \cite{wang2021secure,zhou2024shortcut}}  & $O(n(\kappa + l_v))$ & $O(nN_u(l_k+l_v))$ & $O(n(\kappa + l_v \log n))$ & $O(n(\kappa + l_v \log n))$
        \\
        \hline
        \textbf{Ours ($\protocolsym{aPSI}$-based)} & $O(n(\kappa + \log^2 n))$ & $O(1)^{\bullet}$ & $O(n(\kappa + \log^2 n + l_v \log n))$ & $O(n l_v \log n)$ \\
        \hline
\end{tabular}}
\begin{tablenotes}
  \tiny
  \item $^{\star}$ ShortCut \cite{zhou2024shortcut} does not support PK-FK join, the complexity is considered as the same as CPSI \cite{rindal2021vole} + $\functionsym{trav}$.
  \item $^{\bullet}$\textit{Free} MPC overhead, meaning it does not involve any MPC operations.
\end{tablenotes}
\caption{Communication complexity comparison of different protocols for view generation and refresh. }
\vspace{-0.25in}
\label{tab:view_complexity}
\end{table}

\iffullversion

For two tables $\tablesym{R}^0, \tablesym{R}^1$ with join key $k$, let $\kappa$ be the computational security parameter, $n$ be the size of tables, $N_u$ be the number of update rows, $l_k$ be the bit length of the join key attribute, and $l_v$ be the bit length of the payload attributes of $\tablesym{R}^0$.
We illustrate the communication complexity of our view generation/refresh protocols in Tab.~\ref{tab:view_complexity}.

\noindent
\textbf{Baseline}.
We adopt the protocol in \cite{wang2021secure} that is based on CPSI with payload \cite{rindal2021vole} as our baseline for PK-PK join.
Specifically, PK-PK join requires the parties to execute the CPSI where $\alice$ uses the payload attributes of $\tablesym{R}^0$ as the input payload to CPSI, after which the join result is directly obtained.
Thus, the cost of baseline for view generation and refresh are the same, which is $O(n(\kappa + l_v))$.
For view generation of PK-FK join, additional operations are required to complete the join process as suggested in \cite{han2022scape}, including switching and duplicating payload attributes. 
Specifically, $\bob$, who holds the table with the non-unique join key attributes, should locally sort $\tablesym{R}^1$ ordered by $k$ and permute the shared payload of $\tablesym{R}^0$ by invoking $\protocolsym{osn}^p$, which takes $O(n l_v\log n)$ bits of communication. Then, the parties should invoke $\protocolsym{trav}$ to duplicate the tuples, which takes $O(n l_v)$ bits of communication.
Therefore, the communication cost of the baseline is $O(n(\kappa + l_v\log n))$ bits.
For view refresh of PK-PK join, we set the baseline as ShortCut \cite{zhou2024shortcut}.

\noindent\textbf{Our protocols}.
$\functionsym{cPSI}$ takes $O(n\kappa)$ bits of communication for size-$n$ input.
Furthermore, note that $\functionsym{osn}^s, \functionsym{shuffle}, \functionsym{perm}^\mathsf{s}$ and $\functionsym{invp}^\mathsf{s}$ take communication cost of $O(nl\log n)$ bits for input with $l$-bits. Since $l = \log n$ bits are required to represent a permutation on $n$ input, our $\protocolsym{aPSI}$ for PK-PK join takes $O(n (\kappa + \log^2 n))$ bits of communication. 
The difference between our PK-PK and PK-FK join protocols lies in the switch and duplicate operation of payload attributes. Thus, the communication cost of the PK-FK join protocols increases by $O(nl_v\log n)$ compared to the corresponding PK-PK join protocols.
Overall, our approach for view refreshing reduced the complexity from $O(n(\kappa + l_v))$ ($O(n(\kappa + l_v \log n))$) to $O(1)$ ($O(n l_v \log n)$) for PK-PK (PK-FK) join, respectively. 

\else 

For PK-PK join, we set \cite{wang2021secure} as our baseline, which is to use state-of-the-art (SOTA) CPSI \cite{rindal2021vole} to allow joining with payloads.
For PK-FK join, the baseline protocol requires additional operations, including switching and duplicating payload attributes (see steps 2-3 of our protocol in Appendix \ref{sec:pkfkGenRefresh}), to complete the join process, which is suggested in \cite{han2022scape}.

For two tables $\tablesym{R}^0, \tablesym{R}^1$ with join key $k$, let $\kappa$ be the computational security parameter, $n$ be the size of tables, $N_u$ be the number of update rows, $l_k$ be the bit length of the join key attribute, and $l_v$ be the bit length of the payload attributes of $\tablesym{R}^0$.
We illustrate the communication complexity of our view generation/refresh protocols in Tab.~\ref{tab:view_complexity} and defer the analysis details to the full version.

\fi

\vspace{-0in}
\section{Group Aggregation Protocols}
\label{sec:group}
\vspace{-0.05in}

Thanks to the payload-independence feature of our view, the input payload to GA protocols is plaintext, which enables further optimization of GA. Next, we describe our protocols for securely evaluating the GA query $\mathcal{G}_{(g_0, g_1), \aggfun{aggs}(v_0, v_1)}(\tablesym{R^0} \Join_k \tablesym{R^1})$ that performs on our proposed view $\mathcal{V}$. For $\aggfun{count}$, $v_0$ ($v_1$) is set to a vector of 1s to enable aggregation.
\iffullversion 
The high-level design ideas of our protocols are outlined in Fig. \ref{fig:groupIdea}.
\begin{figure*}[t!]
    \centering
    \includegraphics[width=0.6\linewidth]{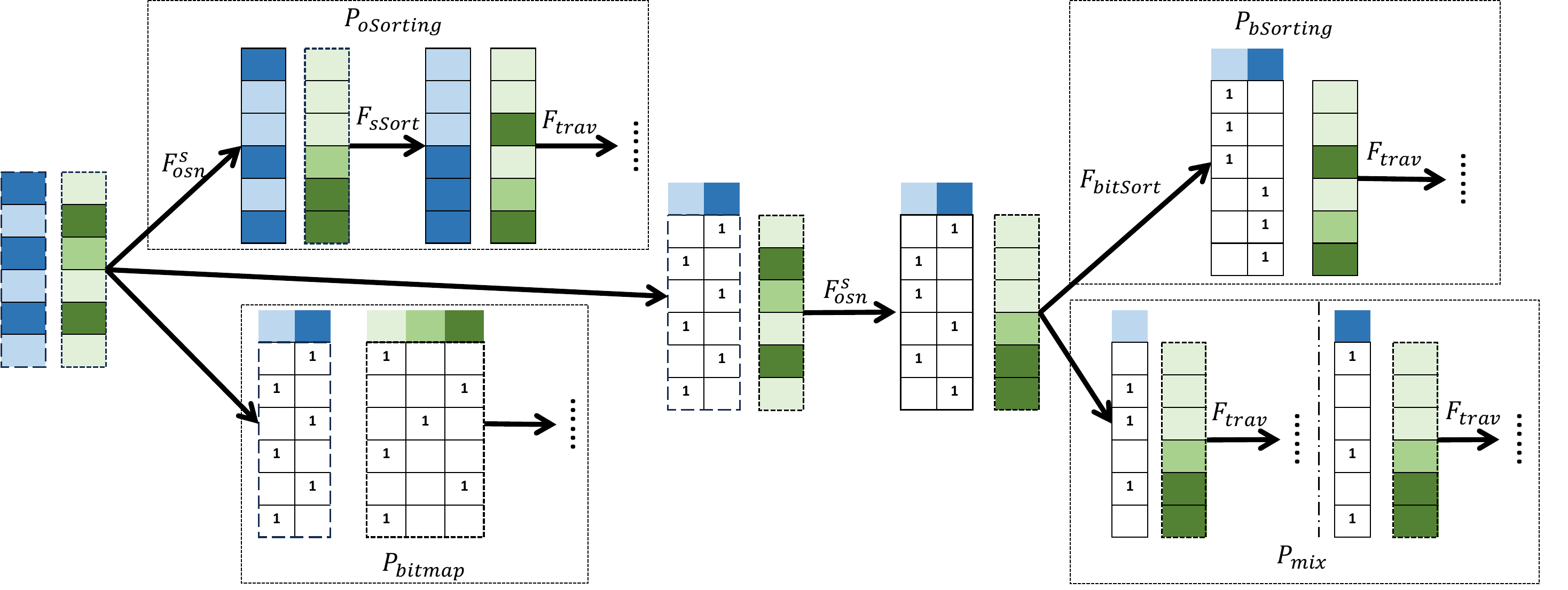}
    \vspace{-0.1in}
    \caption{\small The high-level ideas of GA protocols, where 0s in the bitmap are not displayed.}
    \vspace{-0.1in}
    \label{fig:groupIdea}
\end{figure*}
\else 
Due to space limitations, we introduce only the basic ideas and defer detailed protocols in the full version \cite{peng2024mapcomp}.
\fi

\begin{figure}\small
    \framebox{\begin{minipage}{0.97\linewidth}
        \begin{trivlist} 
            \item \textbf{Input:} 
             The query $\mathcal{G}_{(g_0, g_1), \aggfun{aggs}(v_0, v_1)}(\tablesym{R^0} \Join_k \tablesym{R^1})$. 
            $\alice$ inputs $\mathcal{V}_0 = (\pi_0, \shareb{E}_0, \tablesym{J^0})$. $\bob$ inputs $\mathcal{V}_1 = (\pi_1, \shareb{E}_1, \tablesym{J^1})$, W.L.O.G. we assume $|\pi_0| = |\pi_1| = n$.
            
            \item \textbf{Protocol:} 

            \vspace{-0in}
            
            \begin{enumerate}

                \item $\bob$ generates a length-$m$ permutation $\sigma_b$ such that for $i\in[m-1]$, $\tablesym{J^1_{\sigma_b(i)}}[g_1] \leq \tablesym{J^1_{\sigma_b(i+1)}}[g_1]$.
                
                \item $\bob$ computes and shares $\tablesym{T^{(1)}} = \sigma_b \cdot \tablesym{J^1}$.
                
                \item Invoke $\functionsym{osn}^\mathsf{s}$ and append results into $\tablesym{T^{(1)}}$, where $\bob$ acts as receiver with input $\sigma_b$. $(\shareb{\tablesym{T^{(1)}}[e]}, \share{\tablesym{T^{(1)}}[g_0]}, \share{\tablesym{T^{(1)}}[v_0]}) \leftarrow \functionsym{osn}^\mathsf{s}(\sigma_b, (\shareb{E}, \share{\tablesym{J^0}[g_0]}, \share{\tablesym{J^0}[v_0]}))$.
                
                \item Run stable sorting $(\share{\pi_{g_0}}, (\shareb{\tablesym{T^{(2)}}[e]}, \share{\tablesym{T^{(2)}}[g_0]})) \leftarrow \functionsym{sSort}(\shareb{\tablesym{T^{(1)}}[e]}, \share{\tablesym{T^{(1)}}[g_0]})$.
                
                \item Compute $\share{\tablesym{T^{(2)}}}\leftarrow \functionsym{perm}^s(\share{\pi_{g_0}}, \share{\tablesym{T^{(1)}}})$.

                \item Compute validity flag $F$: $\shareb{f_i} = \shareb{\tablesym{T^{(2)}_{i}}[e]}  \odot (\lnot \functionsym{eq}(\share{\tablesym{T^{(2)}_i}[g_0] || \tablesym{T^{(2)}_i}[g_1]}, \share{\tablesym{T^{(2)}_{i+1}}[g_0] || \tablesym{T^{(2)}_{i+1}}[g_1]}))$.

                \item Init $\share{\tablesym{T^{(3)}}} = \emptyset$ and compute for $u \in \{0, 1\}$: 
                \begin{enumerate}
                    \item $\share{\tablesym{T^{(3)}}[r_u]} \leftarrow \functionsym{trav}(\share{\tablesym{T^{(2)}}[g_0] || \tablesym{T^{(2)}}[g_1]}, \share{\tablesym{T^{(2)}}[v_u]}, \aggfun{agg}_u)$.

                    \item For $i\in [n]$: 
                    $\share{\tablesym{T^{(3)}_i}[g_u]} = \functionsym{mux}(\shareb{f_i}, \share{\tablesym{T^{(2)}_i}[g_u]}, \share{\perp})$, \\ $\share{\tablesym{T^{(3)}_i}[r_u]} = \functionsym{mux}(\shareb{f_i}, \share{\tablesym{T^{(2)}_i}[r_u]}, \share{\perp})$.
                \end{enumerate}

            \item Invoke $\functionsym{osn}^{\mathsf{s}}$ with $\alice$ inputs a random permutation $\alpha$: $\share{\tablesym{T^{(4)}}} \leftarrow \functionsym{osn}^{\mathsf{s}}(\alpha, \share{\tablesym{T^{(3)}}})$.

            \vspace{-0in}
            
            \end{enumerate}
            \item \textbf{Output:} Reveal $\share{\tablesym{T^{(4)}}}$ to $\bob$.
        \end{trivlist}
    \end{minipage}}
    \vspace{-0.1in}
    \caption{(First attempt) Sorting-based GA protocol $\protocolsym{sorting}$.}
    \vspace{-0.1in}
    \label{protocol:sortbasedgrouppre}	
\end{figure}

\vspace{-0.05in}
\subsection{Optimized Sorting-based GA Protocol 
\texorpdfstring{$\protocolsym{oSorting}$}{Lg}} 
\label{sec:sortBased}
\vspace{-0.05in}

A naive sorting-based solution  \cite{han2022scape,liagouris2021secrecy} for the GA query first makes the rows that belong to the same group adjacent through oblivious sorting and then performs oblivious traversal to compute the aggregation result for each group.
Previous work requires sorting over both $g_0, g_1$ and incurs significant overhead.
Our observation is that when the input is plaintext, the input length of the sorting protocol can be reduced by introducing an oblivious stable sorting. 


\vspace{-0in}
\subsubsection{Protocol details}

\begin{figure*}[t!]
    \centering
    \includegraphics[width=0.9\linewidth]{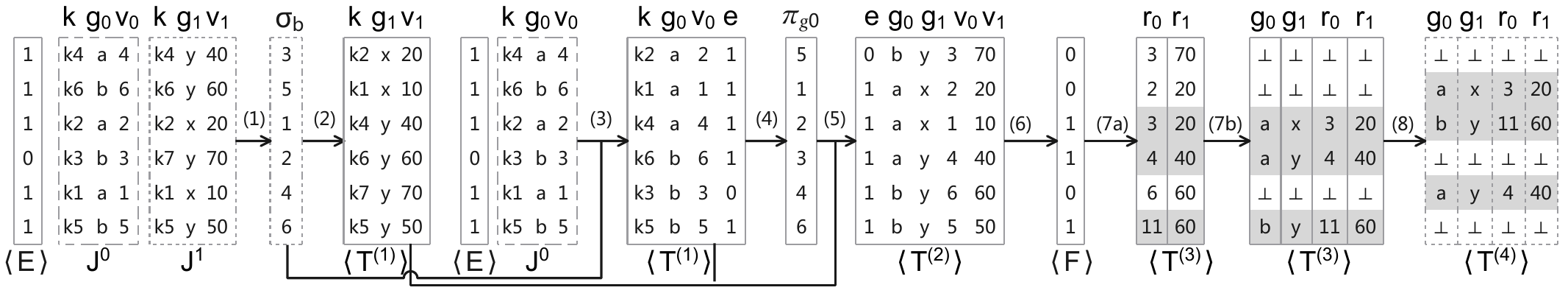}
    \vspace{-0.1in}
    \caption{A running example of $\protocolsym{sorting}$ (Fig. \ref{protocol:sortbasedgrouppre}), where the aggregate functions are $\aggfun{sum}(v_0), \aggfun{max}(v_1)$ and $\domain{g_0} = \{a, b\}$, $\domain{g_1} = \{x, y\}$.}
    \vspace{-0.15in}
    \label{fig:pkpkSortGroup}
\end{figure*}

Since the input of GA is plaintext due to the payload-independence of $\mathcal{V}$, $\bob$ can first locally sort his relation $\tablesym{J^1}$ based on the grouping attribute $g_1$ and obtain the permutation $\sigma_b$. Then, we can obtain $\tablesym{T^1}$ in the order of $g_1$ by applying $\sigma_b$ on $\tablesym{J^0}$ and $\tablesym{J^1}$. Next, parties perform stable sorting $\functionsym{sSort}$ based on $e$ and $g_0$ and invoke $\functionsym{perm}^s$ to obtain $\tablesym{T^{(2)}}$ that is sorted in the lexicographic order of attributes $e, g_0, g_1$. The tuples not belonging to the join result are sorted to the front, and all valid tuples within the same group are sorted together.  
After sorting, the parties compute the validity flag $F$ representing whether the corresponding tuple is the last valid tuple of its group.
Then, in step 7(a), the oblivious traversal is performed on shared values of attributes $v_0, v_1$ to obtain the aggregation result.
After that, $\tablesym{T^{(3)}_i}[r_u]$ with $f_i = 1$ stores the aggregate result over the tuples within the same group.
To hide the number of tuples in each group, $\tablesym{T^{(3)}}$ must be randomly shuffled. Since the result will be revealed to $\bob$, one invocation of $\functionsym{osn}$ with $\alice$ as receiver is enough.
A running example is shown in Fig. \ref{fig:pkpkSortGroup}.

\vspace{-0in}
\subsubsection{Further optimizations}

\iffullversion 

We note that invoking $\functionsym{perm}^s$ is expensive and can be avoided by invoking cheaper $\functionsym{perm}^\mathsf{p}$, which permutes a plain vector based on a shared permutation (described in Appendix \ref{appendix:plainPerm}). 
Then, we optimize $\protocolsym{sorting}$ to obtain the final optimized sorting-based GA protocol $\protocolsym{oSorting}$, and the details are deferred in Appendix \ref{appendix:oSorting}.

In addition, considering that the GA result will be revealed directly, when the aggregation operation is $\aggfun{sum}$/$\aggfun{count}$, the above protocol can be further optimized to avoid prefix operations ($\protocolsym{trav}$) of $O(\log n)$ rounds by the receiver's local operations. See Appendix \ref{appendix:sumCount} for details. 

\else 

We note that invoking $\functionsym{perm}^s$ is expensive and can be avoided by invoking cheaper $\functionsym{perm}^\mathsf{p}$, which permutes a plain vector based on a shared permutation (deferred in the full version). 
Then, we optimize $\protocolsym{sorting}$ to obtain the final optimized sorting-based GA protocol $\protocolsym{oSorting}$.

In addition, considering that the GA result will be revealed directly, when the aggregation operation is $\aggfun{sum}$/$\aggfun{count}$, the above protocol can be further optimized to avoid prefix operations ($\protocolsym{trav}$) of $O(\log n)$ rounds by the receiver's local operations. 

\fi

\vspace{-0in}
\subsubsection{Supporting for One-side Case}

For the special case where the grouping attributes are owned by one party, \eg $\bob$, a single invocation of OSN satisfies to achieve oblivious grouping. Specifically, $\bob$ can locally group his table by locally sorting based on $g_1$ to obtain $\pi$, then permute the remaining payload based on $\pi$ by invoking an OSN. In this case, the oblivious sorting (\eg steps 4-5 in Fig.~\ref{protocol:sortbasedgrouppre}) can be avoided, reducing the communication cost to $O(n\log nl_v^0 + nl_v)$ bits. We denote this protocol as the one-side version of $\protocolsym{oSorting}$.

\vspace{-0in}
\subsection{Bitmap-assisted Sorting-based GA Protocol 
\texorpdfstring{$\protocolsym{bSorting}$}{Lg}}
\label{sec:bitmap_sort}
\vspace{-0in}

\iffullversion

In \S \ref{sec:sortBased}, we improve oblivious sorting with stable sorting of shorter input based on the fact that $g_1$ is plaintext. We observed that when $g_0$ is also plaintext, we can exploit the input cardinality to further optimize the oblivious stable sorting. Inspired by Radix sorting \cite{asharov2022efficient}, we first design a more efficient stable sorting protocol $\protocolsym{bitSort}$ for the cases where input $g_0$ is plaintext and has a small cardinality by utilizing a bitmap, which may be of independent interest. The idea is to first encode $g_0$ as a bitmap, the bit-vector of which indicates which bucket the corresponding tuple should be placed in. Then, the result of sorting can be obtained by counting and indexing over the bitmap. Since the size of a bitmap is linear with the cardinality of group attribute $d_0 = |\domain{g_0}|$, it is more lightweight compared with the oblivious sort for small $d_0$ input. We defer the details of $\protocolsym{bitSort}$ in Appendix \ref{appendix:bitSort}.

Based on $\protocolsym{bitSort}$, we present the bitmap-assisted sorting-based GA protocol $\protocolsym{bSorting}$ in Fig.~\ref{protocol:mixedBitmapSort}. The main difference compared to $\protocolsym{oSorting}$ is replacing $\functionsym{sSort}$ with $\protocolsym{bitSort}$ to obtain better efficiency. The first two steps aim to generate bitmap encoding for tuples in the join result, such that $\tablesym{J^0_i}[b^j] = 1$ iff $\tablesym{J^0_i}$ belongs to the join result and the group value of $\tablesym{J^0_i}$ equals $\domain{g_0}_j$. Thus, the result permutation in step 4 will sort all tuples in the lexicographic order of attributes $e, g_0, g_1$. The following steps 5-7 permute relations based on $\sigma_b$ and $\sigma_{g_0}$, similar to $\protocolsym{oSorting}$. Thus, oblivious grouping based on $g_0,g_1$ is achieved. The protocol $\protocolsym{invp}^\mathsf{p}(\share{\pi}, X)$ is deferred in Appendix \ref{appendix:plainPerm}.

\begin{figure}\small
    \framebox{\begin{minipage}{0.97\linewidth}
        \begin{trivlist} 
        
            \item \textbf{Input:} 
            Same as Fig.~\ref{protocol:sortbasedgrouppre}.
            
            \item \textbf{Protocol:} After step 2 of Fig.~\ref{protocol:sortbasedgrouppre}:
            \begin{enumerate}
                \item $\alice$ generates bitmap encoding of $g_0$: $\{\tablesym{J^0}[b^1], \dots, \tablesym{J^0}[b^{d_0}]\}$.
                
                \item Compute for $j \in [d_0], i\in [n]$: $\shareb{\tablesym{J^0_i}[b^j]} = \shareb{e_i} \odot \shareb{\tablesym{J^0_i}[b^j]}$.
                
                \item Invoke $\functionsym{osn}^\mathsf{s}$ and append results into $\tablesym{T^{(0)}}$, where $\bob$ acts as receiver with input $\sigma_b$. $(\shareb{\tablesym{T^{(0)}}[b^1]}, \dots, \shareb{\tablesym{T^{(0)}}[b^{d_1}]})  \leftarrow \functionsym{osn}^\mathsf{s}(\sigma_b, (\shareb{\tablesym{J^0}[b^1]}, \dots, \shareb{\tablesym{J^0}[b^{d_0}]}))$.
            
                \item $\share{\pi_{g_0}} \leftarrow \functionsym{bitSort}(\shareb{\tablesym{T^{(0)}_i}[b^1]},  \dots, \shareb{\tablesym{T^{(0)}_i}[b^{d_0}]})$.
            
                \item The parties invoke $\functionsym{invp}^\mathsf{p}$ where $\bob$ acts as sender, and append results into $\tablesym{T^{(2)}}$: $(\share{\tablesym{T^{(2)}}[g_1]}, \share{\tablesym{T^{(2)}}[v_1]}, \share{\rho})\leftarrow \functionsym{invp}^\mathsf{p}(\share{\pi_{g_0}}, \tablesym{T^{(1)}}[g_1], \tablesym{T^{(1)}}[v_1], \sigma_b)$.
            
                \item The parties invoke $\functionsym{perm}^\mathsf{p}$ and append results into $\tablesym{T^{(2)}}$, where $\alice$ acts as sender: $(\share{\tablesym{T^{(2)}}[g_0]}, \share{\tablesym{T^{(2)}}[v_0]}) \leftarrow \functionsym{perm}^\mathsf{p}(\share{\rho},  (\tablesym{J^0}[g_0], \tablesym{J^0}[v_0]))$.
            
                \item $\share{\tablesym{T^{(2)}}[e]} \leftarrow \functionsym{perm}^\mathsf{s}(\share{\rho}, \shareb{E})$.
            \end{enumerate}
            
            \item \textbf{Then:} Run the remainder after step 5 in Fig.~\ref{protocol:sortbasedgrouppre}.
        \end{trivlist}
    \end{minipage}}
    \vspace{-0.1in}
    \caption{Bitmap-assisted sorting-based GA protocol $\protocolsym{bSorting}$.}
    \vspace{-0.15in}
    \label{protocol:mixedBitmapSort}	
\end{figure}

\else

In \S \ref{sec:sortBased}, we improve oblivious sorting with stable sorting of shorter input based on the fact that $g_1$ is plaintext. We observed that when $g_0$ is also plaintext, we can exploit the input cardinality to further optimize the oblivious stable sorting. 
The idea is to first encode $g_0$ as a bitmap, the bit-vector of which indicates which bucket the corresponding tuple should be placed in. Then, the result of sorting can be obtained by counting and indexing over the bitmap. Since the size of a bitmap is linear with the cardinality of group attribute $d_0 = |\domain{g_0}|$, it is more lightweight compared with the oblivious sort for small $d_0$ input. 

\fi

\vspace{-0.0in}
\subsection{Mixed GA Protocol \texorpdfstring{$\protocolsym{mix}$}{Lg}}
\label{sec:mixedPto}
\vspace{-0.0in}

\iffullversion

Instead of dividing groups based on oblivious sorting, we observe that the grouping can also be achieved by using a bitmap since each bit-vector of the bitmap naturally specifies a distinct group value, based on which we propose our mixed GA protocol $\protocolsym{mix}$. The high-level idea is to obtain the aggregation result for each group divided by $g_0, g_1$, then divide groups based on $g_1$ by local sorting and invoking $\functionsym{osn}^\mathsf{s}$ (the same as $\protocolsym{oSorting}$).

Now we present the details of $\protocolsym{mix}$ in Fig. \ref{protocol:mixedGroup}. First, $\alice$ generates the bitmap of $g_0$, where each bit-vector represents whether the input tuples belong to a specific group of $g_0$. After step 2, $\bob$ locally sorts $\tablesym{J^1}$ based on $g_1$, and computes the group indicator $C$ representing the boundary of groups divided by $g_1$. 
Then, $\functionsym{mux}$ is invoked to set the values that do not belong to the current group of $g_0$ as $\perp$ to eliminate their effect. Finally, the aggregation result of groups based on $g_0,g_1$ can be obtained by invoking $\functionsym{trav}$. Such that for $i\in[n]$ and $j\in[d_0]$,  $\tablesym{T^{(2)}_i}[r_u^j]$ stores the aggregation result of group $ \domain{g_0}_j||\tablesym{T^{(1)}_i}[g_1]$ if $c_{i} = 1$, and $\tablesym{T^{(2)}_i}[r_u^j] = \perp$ otherwise. A running example of $\protocolsym{mix}$ is shown in Fig.~\ref{fig:groupMixed}. $\protocolsym{mix}$ mixes the use of bitmap (for $g_0$) and local sorting (for $g_1$) and avoid costly oblivious sorting. As a trade-off, the $\functionsym{trav}$ should be invoked for $d_0$ times for each distinct group of $g_0$. 

\begin{figure}\small
    \framebox{\begin{minipage}{0.97\linewidth}
        \begin{trivlist} 
        
            \item \textbf{Input:} 
            Same as Fig.~\ref{protocol:sortbasedgrouppre}.
            
            \item \textbf{Protocol:} 
            \begin{enumerate}
                \item $\alice$ generates bitmaps of $g_0$ in $\tablesym{J^0}$: $\{\tablesym{J^0}[b^1], \dots, \tablesym{J^0}[b^{d_0}]\}$. 

                \item The parties run steps 1-3 in Fig.~\ref{protocol:sortbasedgrouppre}, and replace $\tablesym{J^0}[g_0]$ with $\{\tablesym{J^0}[b^1], \dots, \tablesym{J^0}[b^{d_0}]\}$ in step 3.

                \item $\bob$ locally computes and shares the group indicator $C$. $\forall i\in [n]$: $c_i = ind(\tablesym{T^1_{i}}[g_1] \neq \tablesym{T^1_{i + 1}}[g_1])$.

                \item The parties initialize $\share{\tablesym{T^2}} = \emptyset$, and process in parallel for $u \in \{0,1\}, j \in [d_0]$:
                \begin{enumerate}
                    \item The parties compute a temporal vectors $A$ for $i\in [n]$: $\share{a_i} = \functionsym{mux}(\shareb{\tablesym{T^{(1)}_i}[e]} \odot \shareb{\tablesym{T^{(1)}_i}[b^j]}, \share{\tablesym{T^{(1)}_i}[v_u]}, \share{\perp})$. 

                    \item $\share{\tablesym{T^{(2)}}[r_u^j]} \leftarrow \functionsym{trav}(\share{\tablesym{T^{(1)}}[g_1]}, \share{A}, \aggfun{agg}_u)$.
                    
                    \item For $i\in[n]$: $\share{\tablesym{T^{(2)}_i}[r^j_u]} = \functionsym{mux}({\shareb{ c_{i}}, \share{\tablesym{T^{(2)}_i}[r_u^j]}, \share{\perp}})$.

                \end{enumerate}
            \end{enumerate}
            \item \textbf{Output:} Reveal $\share{\tablesym{T^2}}$ to $\bob$.
        \end{trivlist}
    \end{minipage}}
    \vspace{-0.05in}
    \caption{Mixed GA protocol $\protocolsym{mix}$.}
    \vspace{-0.05in}
    \label{protocol:mixedGroup}	
\end{figure}

\begin{figure}[t!]
\centering
\includegraphics[width=0.7\linewidth]{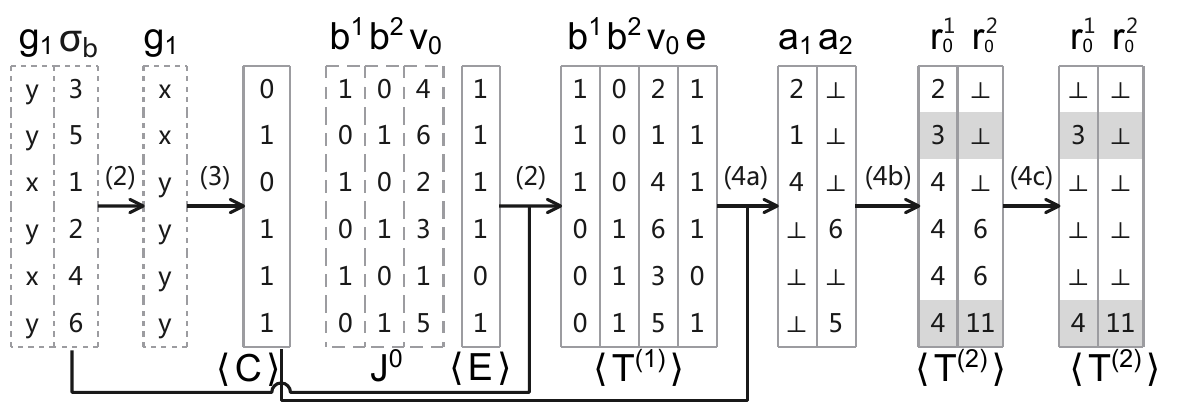}
\vspace{-0.1in}
\caption{\small Partially executed example of the $\protocolsym{mix}$, where the aggregate function is $sum(v_0)$, and the numbers in the dark blocks are results. The input data are the same as Fig.\ref{fig:pkpkSortGroup}.}
\vspace{-0.2in}
\label{fig:groupMixed}
\end{figure}

\else

Instead of dividing groups based on oblivious sorting, we observe that the grouping can also be achieved by using a bitmap since each bit-vector of the bitmap naturally specifies a distinct group value. To obtain the aggregation result for each group divided by $g_0, g_1$, $\protocolsym{mix}$ first divide tuples into groups based on $g_0$ by encoding $g_0$ as a bitmap, and then divide groups based on $g_1$ by local sorting and invoking $\functionsym{osn}^\mathsf{s}$ (the same as $\protocolsym{oSorting}$). $\protocolsym{mix}$ mixes the use of bitmap (for $g_0$) and local sorting (for $g_1$) and avoid costly oblivious sorting. As a trade-off, the $\functionsym{trav}$ should be invoked for $d_0$ times for each distinct group of $g_0$. 

\fi

\vspace{-0.05in}
\subsection{Bitmap-based GA Protocol \texorpdfstring{$\protocolsym{bitmap}$}{Lg}}
\vspace{-0.0in}

\iffullversion

We propose $\protocolsym{bitmap}$ to optimize GA for cases where $d_0d_1$ is small by encoding both  $g_0,g_1$ as bitmaps. The high-level idea is to divide distinct groups of $g_0,g_1$ via the bit-vectors of bitmaps, so any oblivious sorting or functions that require OSN (such as $\functionsym{shuffle}$ or $\functionsym{perm}^s$) are avoided. The cost is additional aggregation processes of $d_0d_1$ times, so it is more efficient in small $d_0d_1$ cases, which we confirm in \S\ref{sec:evaluation}. 

\begin{figure}\small
    \framebox{\begin{minipage}{0.97\linewidth}
        \begin{trivlist}
            \item \textbf{Input:} 
            Same as Fig.~\ref{protocol:sortbasedgrouppre}.
            
            \item \textbf{Protocol:} 
            \begin{enumerate}
                \item $\alice$ generates $d_0$ bit vectors representing the bitmap encoding of $g_0$ in $\tablesym{J^0}$: $\{\tablesym{J^0}[b_0^1], \dots, \tablesym{J^0}[b_0^{d_0}]\}$. Similarly, $\bob$ generates bitmap encoding of $g_1$ in $\tablesym{J^1}$: $\{\tablesym{J^1}[b_1^1], \dots, \tablesym{J^1}[b_1^{d_1}]\}$. 
                
                \item Initialize $\share{\tablesym{T}} = \emptyset$. For $j\in [d_0], p\in [d_1]$, process as follow:
                \begin{enumerate}
                    \item For $i\in[n]$, compute temporary shared vectors $Q^0, Q^1$: 

                    \begin{itemize}
                        \item $\share{q^1_i} = \functionsym{mul}(\tablesym{J^0_i}[b_0^j], \tablesym{J^1_i}[b_1^p] \cdot \tablesym{J^1_i}[v_1])$.
                        \item $\share{q^0_i} = \functionsym{mul}(\tablesym{J^1_i}[b_1^p], \tablesym{J^0_i}[b_0^j] \cdot \tablesym{J^0_i}[v_0])$.
                    \end{itemize}
                    
                    \item Compute and append $(\domain{g_0}_j, \domain{g_1}_p, \aggfun{agg}_0(\share{Q^0}, \shareb{E}),$ $\aggfun{agg}_1(\share{Q^1}, \shareb{E}))$ to result table $\share{\tablesym{T}}$.
                \end{enumerate}
            \end{enumerate}
            \item \textbf{Output:} Reveal $\share{\tablesym{T}}$ to $\bob$.
        \end{trivlist}
    \end{minipage}}
    \vspace{-0.1in}
    \caption{Bitmap-based GA protocol $\protocolsym{bitmap}$.}
    \vspace{-0.2in}
    \label{protocol:group4Pk}	
\end{figure}

The details of $\protocolsym{bitmap}$ are illustrated in Fig.~\ref{protocol:group4Pk}. Parties both use bitmaps to divide their own groups. Then, for the tuples that do not belong to a group, they are obliviously set as dummy by invoking $\functionsym{mul}$ with $\tablesym{J_i}[b]$. Aggregations $\aggfun{agg}$ are performed to obtain final results. The round complexity of  $\protocolsym{bitmap}$ depends on the type of aggregation. When $\aggfun{agg}$ is $\aggfun{sum}$ or $\aggfun{count}$, the aggregation is $\sum_{i=1}^n{\functionsym{mux}(\shareb{e_i}, \share{q_i}, \share{0})}$ and can be computed in one round. When $\aggfun{agg}$ is $\aggfun{max}$ or $\aggfun{min}$, the computation can be performed via binary-tree-based method \cite{knott2021crypten,zhou2023bicoptor} that simply invokes $O(n)$ times of comparison in $O(\log n)$ rounds.

\else

We propose $\protocolsym{bitmap}$ to optimize GA for cases where $d_0d_1$ is small by encoding both $g_0,g_1$ as bitmaps. The idea is to divide distinct groups of $g_0,g_1$ via the bit-vectors of bitmaps, so any oblivious sorting or functions that require OSN (such as $\functionsym{shuffle}$ or $\functionsym{perm}^s$) are totally avoided. The cost is additional aggregation processes of $d_0d_1$ times, so it is more efficient in small $d_0d_1$ cases. 

\fi

\vspace{-0.05in}
\subsection{Complexity Analysis}
\label{sec:analysis}
\vspace{-0.05in}

Let $n$ be the size of the view $\mathcal{V}$,  $l_v^0, l_v^1$ ($l_g^0, l_g^1$) be the bit length of values of aggregation attributes $v_0, v_1$ (grouping keys $g_0, g_1$) respectively. $l_v = l_v^0 + l_v^1, l_g = l_g^0 + l_g^1, l = l_v + l_g$. We illustrate the communication complexity of our GA protocols in Tab.~\ref{tab:groupComp}.
We set the sorting-based approach with secret-shared input $g_0, g_1$ as the baseline, which captures the GA process of \cite{poddar2021senate,wang2021secure,liagouris2021secrecy,han2022scape}.

\iffullversion

For the baseline and our $\protocolsym{oSorting}$, oblivious sorting is the most expensive process, and the main difference between the cost of baseline and our $\protocolsym{oSorting}$ comes from the input bit length of the oblivious sorting algorithm.
Specifically, to sort $n$ $l$-bit secret shared data, oblivious bitonic sorting takes the communication cost of $O(n\log^2 n(l + \log n))$ bits and $O(\log^2 n\log (l + \log n))$ rounds, oblivious quick sorting takes the communication cost of $O(n\log n(l + \log n))$ bits and $O(\log n\log (l + \log n))$ rounds.
Thus, if $\protocolsym{oSorting}$ is instanced with bitonic sorting, it will take $O(n\log n(l_g^1 \log n + l + \log^2 n))$ bits of communication; if using quick sorting, it will take $O(n\log n(l_g^1 + l + \log n))$ bits of communication.

In $\protocolsym{bSorting}$, $\protocolsym{bitSort}$ takes $O(d_0n\log n)$ bits and $O(1)$ rounds.
$\functionsym{perm}^\mathsf{p}$ and $\functionsym{invp}^\mathsf{p}$ takes $O(1)$ rounds and $O(nk\log n)$ bits communication for $k$-bit input. 
For $\aggfun{max/min}$, $\functionsym{trav}$ takes $O(\log n \log l_v)$ rounds and $O(nl_v)$ bits of communication. 
Therefore, the communication round of $\protocolsym{bSorting}$ is dominated by $\functionsym{trav}$, and its communication cost is $O(n\log n(d_0 + l + \log n))$ bits.

For $\protocolsym{mix}$, it invokes $\functionsym{osn}^{\mathsf{p}}$ once, and $\functionsym{trav}$ and $\functionsym{mux}$ for $O(d_0)$ times, resulting the cost of $O(n\log n(d_0 + l_v^0) + nd_0l_v)$ bits. The round complexity is also dominated by the aggregate functions. For $\protocolsym{bitmap}$, the communication grows linearly with $d_0$ and $d_1$, and the aggregation (step 2) can be computed in parallel, which takes $O(1)$ rounds for $\aggfun{sum}$/$\aggfun{count}$ and $O(\log n\log l_v)$ rounds for $\aggfun{max}$/$\aggfun{min}$.

\fi

\begin{table}[t]
    \centering  
    \small
    \renewcommand{\arraystretch}{1.2}
    \resizebox{0.8\columnwidth}{!}{
        \begin{tabular}{| c | c | c |}
            \hline
            \textbf{Pto.} & \textbf{Comm. cost (bits)} & \textbf{Comm. Rounds} \\
            \hline
            Baseline & $O(n\log n(l_g + l + \log n))$ & $O(\log n\log (l_g + \log n))$ \\
            \hline
            $\protocolsym{oSorting}$ & $O(n\log n(l_g^1 + l + \log n))$ & $O(\log n\log (l_g^1 + \log n))$ \\
            \hline
            $\protocolsym{bSorting}$ & $O(n\log n(d_0 + l + \log n))$ & $O(1)$ / $O(\log n\log l_v)$ \\
            \hline
            $\protocolsym{mix}$ & $O(n\log n(d_0 + l_v^0) + nd_0l_v)$ & $O(1)$ / $O(\log n\log l_v)$ \\
            \hline
            $\protocolsym{bitmap}$ & $O(d_0d_1nl_v)$ & $O(1)$ / $O(\log n\log l_v)$ \\
            \hline
        \end{tabular}
    }
    \caption{Complexity analysis of GA protocols. The communication rounds of some protocols depend on $\aggfun{agg}$, which are $O(1)$ for $\aggfun{sum}$/$\aggfun{count}$ and $O(\log n\log l_v)$ for $\aggfun{max}$/$\aggfun{min}$.}
\label{tab:groupComp}
\vspace{-0.3in}
\end{table}

\vspace{-0.0in}
\section{Implementation and Evaluation}
\label{sec:evaluation}
\vspace{-0.05in}
We demonstrate the evaluation results by addressing the following research questions (RQ):

\vspace{-0.0in}
\begin{itemize}
    \item \textbf{RQ1:} Does our proposed materialized view offer efficiency advantages of view operations over existing non-view baseline? (\S \ref{evaluation:view})
    \item \textbf{RQ2:} Does our proposed GA protocols outperform the traditional sorting-based baseline?  Which GA protocol should one use in a given setting? (\S \ref{evaluation:GA})
    \item \textbf{RQ3:} Does MapComp scale to large-scale datasets and real-world queries? To what extent can MapComp enhance over the non-view JGA baseline? (\S \ref{eva:real-world})
\end{itemize}

\vspace{-0.05in}
\subsection{Experimental Setups}
\vspace{-0.05in}

We implement the prototype of \oursys based on Java with JDK16. 
Our code is publicly available at \url{https://github.com/jerryp1/mapcomp}. 
We evaluate our protocols on two physical machines with Intel$^\circledR$ Core$^{\text{TM}}$ i9-9900K 3.60GHz CPU and 128GB RAM. The number of available threads for each party is 15.
The network settings include LAN (2.5Gbps bandwidth) and WAN (100Mbps bandwidth with 40ms RTT latency). We precompute the multiplication triples used in MPC offline and include this portion of time in the total time. The size of join keys and aggregation attributes is fixed at 64 bits. For GA protocols, we evaluate the most time-consuming case where the group attributes come from two parties.

\vspace{-0in}
\subsection{Performance Gain of View Operations}
\label{evaluation:view}
\vspace{-0in}
We address RQ1 by evaluating MapComp's view generation and refresh performance against baselines.
For PK-PK join, we use the protocol in \cite{wang2021secure} that is based on CPSI with payload \cite{rindal2021vole} and Shortcut \cite{zhou2024shortcut} as the baselines for view generation and refresh, respectively. 
The baseline of PK-FK join is set as CPSI with additional oblivious payload duplication via $\protocolsym{trav}$, which is suggested in \cite{han2022scape} in the three-party setting, and we instantiate it in the two-party setting. We ran the experiment in the LAN setting with the databases containing $2^{20}$ tuples. The performance of Shortcut depends on the size of update tuples, and we present the execution time for various update sizes. The results are summarized in Tab.~\ref{tab:view}. Since view generation is a one-time task (\eg can be pre-computed offline) and refresh is more frequently invoked in online tasks, our focus is on the efficiency of view refresh. 

We observe that our protocols significantly outperform the baseline in view refresh. For the PK-PK join, our protocols generate a completely reusable join view. Compared with Shortcut, which takes around 854s\footnote{For $N_u$ update rows, Shortcut \cite{zhou2024shortcut} requires an $O(nN_u)$ comparison that demands $O(nkN_u)$ multiplication triples, while generating triples is heavy in MPC. Therefore, it is more efficient to re-generate the view when there is an update for Shortcut.}. To update the view after one update, the only overhead of our approach is memory IO to access and modify existing tuples, making it extremely efficient with less than 1 millisecond time, and we ignore it in our statistics. For the PK-FK join view, our approaches are partially refresh-free with little MPC operation, \eg $\functionsym{osn}$ and $\functionsym{trav}$, providing a performance edge of up to $13.9 \times$.

\begin{table}[t]
    \small
    \renewcommand{\arraystretch}{1.2}
    \centering
        \resizebox{1\columnwidth}{!}{
    \begin{tabular}{|c|c|| c|c|c|| c|c| c|c| c|c|}
        \hline
        \multicolumn{2}{|c||} {\textbf{View Operations}} & \multicolumn{3}{c||}{ \textbf{View Generation}} & \multicolumn{6}{c|} { \textbf{View Refresh}} \\
        \hline
        \multicolumn{2}{|c||} {\textbf{Len. of Payload(bits)}} & $\mathbf{2^4}$ & $\mathbf{2^6}$ & $\mathbf{2^8}$ & \multicolumn{2}{c|}{$\mathbf{2^4}$} & \multicolumn{2}{c|}{$\mathbf{2^6}$} & \multicolumn{2}{c|}{$\mathbf{2^8}$} \\
        \hline
        \multicolumn{2}{|c||} {\textbf{ Num. of Update Rows } ($N_u$)} & \multicolumn{3}{c||}{-} & 1 & 2 & 1 & 2 & 1 & 2 \\
        \hline
        \multirow{3}*{\textbf{PK-PK}} & Baseline \cite{wang2021secure} 
        & 270.5 & 271.7 & 272.3  & \multicolumn{2}{c|}{270.5}  & \multicolumn{2}{c|}{271.7} & \multicolumn{2}{c|}{272.3} \\
        \cline{2-11}
        & Baseline \cite{zhou2024shortcut} & 271.4 & 265.8 & 272.8  & 854.5 & 1708.7  & 854.6 & 1708.8 & 854.7 & 1706.1 \\
        \cline{2-11}
        ~ & \textbf{Ours ($\protocolsym{secV}$)} & 962.2& 962.2& 962.2& \multicolumn{2}{c|}{\textcolor{mygreen}{0}} & \multicolumn{2}{c|}{\textcolor{mygreen}{0}} & \multicolumn{2}{c|}{\textcolor{mygreen}{0}} \\
        \hline
        \hline
        \multirow{2}*{\textbf{PK-FK}} & Baseline (CPSI \cite{rindal2021vole} + $\functionsym{trav}$) & 302.2& 315.6& 364.9& \multicolumn{2}{c|}{302.1} & \multicolumn{2}{c|}{314.2} & \multicolumn{2}{c|}{363.2} \\
        \cline{2-11}
        ~ & \textbf{Ours ($\protocolsym{secV}$ + $\functionsym{trav}$)} & 944.2& 953.4& 1,000.9& \multicolumn{2}{c|}{\textcolor{mygreen}{21.6}}  & \multicolumn{2}{c|}{\textcolor{mygreen}{28.5}}  & \multicolumn{2}{c|}{\textcolor{mygreen}{76.6}}\\
        \hline
\end{tabular}
}
\caption{Execution time (s) of view operations with input size $n = 2^{20}$ in LAN setting.}
\label{tab:view}
\vspace{-0.2in}
\end{table}

\vspace{-0.0in}
\subsection{Efficiency Comparison of GA Protocols}
\label{evaluation:GA}
\vspace{-0.0in}

We address RQ2 by evaluating our GA protocols over databases with different data sizes and various group cardinalities. We set the widely used oblivious sorting-based approach \cite{poddar2021senate,wang2021secure,liagouris2021secrecy,han2022scape} with secret-shared input $g_0,g_1$ as the baseline. 
The major difference between the baseline and $\protocolsym{oSorting}$ is that the sorting in the baseline trivially requires inputs from \textit{both} parties, while $\protocolsym{oSorting}$ only needs input from a single party. Since sorting is the major efficiency bottleneck, it leads to significant efficiency gaps between the baseline and our protocols. 
We demonstrate the results of \textit{equal} group cardinality cases as below, where $d_0=d_1$ and ranges from $2^2$ to $2^8$.

\begin{figure*}[t!]
    \centering
    \includegraphics[width=0.75\linewidth]{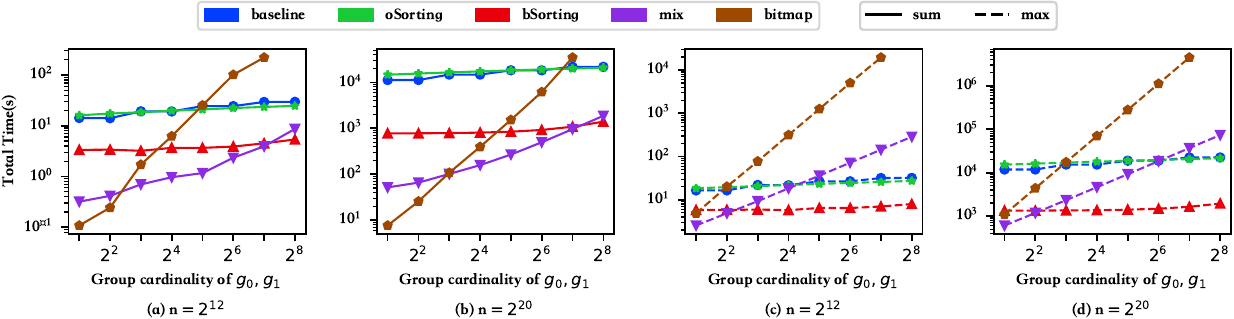}
    \vspace{-0.1in}
    \caption{Running time of GA protocols for equal group cardinality in LAN setting.}
    \vspace{-0.15in}
    \label{fig:smallSizes}
\end{figure*}

\begin{figure*}[t!]
    \centering
    \includegraphics[width=0.75\linewidth]{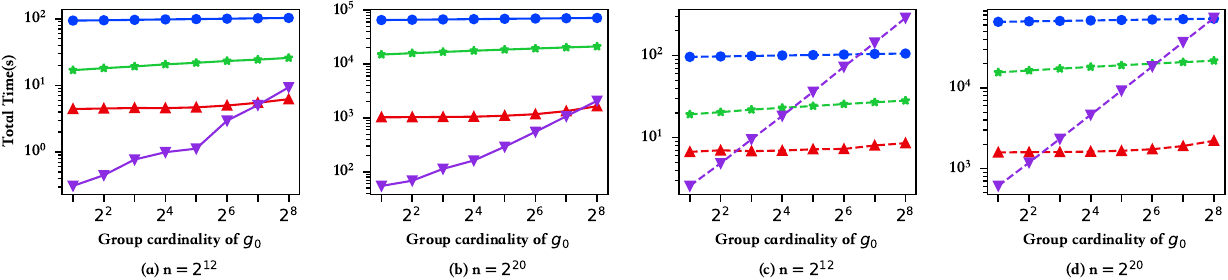}
    \vspace{-0.1in}
    \caption{Running time of GA protocols for unequal group cardinality in LAN setting, where $|\domain{g_1}| = 2^{64}$.}
    \vspace{-0.15in}
    \label{fig:ubSizes}
\end{figure*} 

As shown in Fig.~\ref{fig:smallSizes}, our best approach is $3.5\times \sim 1140.5\times$ faster than the baseline. 
Specifically, when the group cardinality $d=|\domain{g}|$ is small (\eg $2^2$), $\protocolsym{bitmap}$ achieves the best efficiency for group-by-sum due to its total avoidance of oblivious sorting. As the cardinality increases, the linear overhead of $\protocolsym{bitmap}$ ($O(d_0 d_1)$) dominates the cost, and $\protocolsym{mix}$ performs the best after $2^3$. For the comparison of $\protocolsym{mix}$ and $\protocolsym{bSorting}$, as shown in Tab.~\ref{tab:groupComp}, the complexity related to $d_0$ are $n\log n$ and $n(\log n + l_v)$ respectively, which means that the overhead of the $\protocolsym{mix}$ increases faster as $d_0$ increases. Thus, $\protocolsym{bSorting}$ shows better efficiency after $d=2^7$. For group-by-max, the aggregation phase relies on a heavy pairwise comparison \cite{rathee2020cryptflow2} and dominates the cost. Thus, the protocols that require a linear time aggregation ($\protocolsym{bitmap}$ and $\protocolsym{mix}$) have no advantage when $d$ is larger than $2^2$. In this case, our $\protocolsym{bSorting}$ outperforms others. This is because the size of inputs and the efficiency of the sorting algorithm are the main factors impacting the running time for these sorting-based protocols. The superior performance of $\protocolsym{oSorting}$ over the baseline is attributed to the reduced size of sorting input that our view brings, and the advantage over $\protocolsym{bSorting}$ is due to the improvement of our $\protocolsym{bitSort}$. 

\iffullversion

A more obvious observation can be learned from the \textit{unequal} group cardinality cases. 
Since the efficiency of our protocol relies more on the bit-length of the party with smaller group cardinality (the other side can benefits from accelerated processing through local sorting or bitmap encoding), our methods show a more pronounced superiority. 
We compare the efficiency of different GA protocols under \textit{unequal} group cardinality in Fig.~\ref{fig:ubSizes}, where $d_1$ is fixed to $2^{64}$ and $d_0$ varies. 
As shown in Fig.~\ref{fig:ubSizes}, it is demonstrated that $\protocolsym{bSorting}$ and $\protocolsym{oSorting}$ are much more efficient than the baseline in all cases, up to $29.3\times$ improvements. This is because the oblivious sorting with relatively large input ($d_1 = 2^{64}$) is optimized by local sorting. As the complexity of $\protocolsym{bSorting}$ is also linear with $d_0$, $\protocolsym{oSorting}$ is more efficient when $d_0$ is large. For the comparison of the baseline and $\protocolsym{oSorting}$ (regardless of the underlying sorting algorithm), we note the bit length of input for the sorting process of them are $l_{g_0} + l_{g_1}$ and $\mathsf{min}({l_{g_0}, l_{g_1}}) + \log n$ respectively. Thus, $\protocolsym{oSorting}$ outperforms in the cases where $\mathsf{max}({l_{g_0}, l_{g_1}})$ is relatively large when $n$ is fixed, as confirmed in Fig.~\ref{fig:ubSizes}.

\fi

Next, based on the complexity analysis and evaluation results, we answer the second part of RQ2. That is, which GA protocol performs best in a given setting? Our analysis gives the following guidelines for representative dataset size ($2^{20}$) in the LAN setting.

\vspace{-0.05in}
\begin{itemize}
    \item When $d_0,d_1$ are both smaller than $2^3$ and aggregation function is $\aggfun{sum}$, $\protocolsym{bitmap}$ is the best among all approaches.
    \item Apart from the above case, when $d_0$ or $d_1$ is less than $2^7$ and  aggregation function is $\aggfun{sum}$, one should use $\protocolsym{mix}$.
    \item Apart from the above cases, when $d_0$ or $d_1$ is
    less than $2^{12}$, we should use $\protocolsym{bSorting}$. 
    \item When $d_0$ and $d_1$ are both large than $2^{12}$, we should choose $\protocolsym{oSorting}$. 
\end{itemize}

\vspace{-0in}
\subsection{Real-world Queries Simulation}
\label{eva:real-world}
\vspace{-0.0in}

To answer RQ3, we simulated the real-world applications including 4 queries from the TPC-H benchmark and 2 queries on CPDB and TPC-ds.

\partitle{Queries} We use 4 real-world JGA queries from the widely adopted TPC-H benchmark \cite{TPC} to evaluate our framework.
Since the performance of our secure protocols is highly relevant to the dataset size and bit-length of values, the dataset is generated randomly under the specified experiment settings to enable fine-grained observations. 
We assume tables ``customer'' and ``orders'' are from one party, ``lineitem'' from the other party, and ``nation'' is a public table.

We further use two real-world datasets Chicago Police Data (CPDB) \cite{chicago} and TPC-ds \cite{TPC} with chosen queries. For CPDB, we selected two tables Award (807,597 rows) and Complaint (263,315 rows). The query is to count the number of awards received by officers with complaints over a threshold NUM, grouped by the time and award type. For 1GB TPC-ds, we selected Inventory (45000 rows for 1 Day) and web\_sale (983 rows for 1 Day). The query is to sum unsold products per warehouse and product category for a specified date. We assume that two tables in each dataset belong to different parties. 
\iffullversion
More details on query configuration and SQL statements can be found in Appendix \ref{appendix:query}.
\else 
More details on query configuration and SQL statements can be found in our full version \cite{peng2024mapcomp}.
\fi

\begin{figure}[t!]
    \centering
    \includegraphics[width=0.7\linewidth]{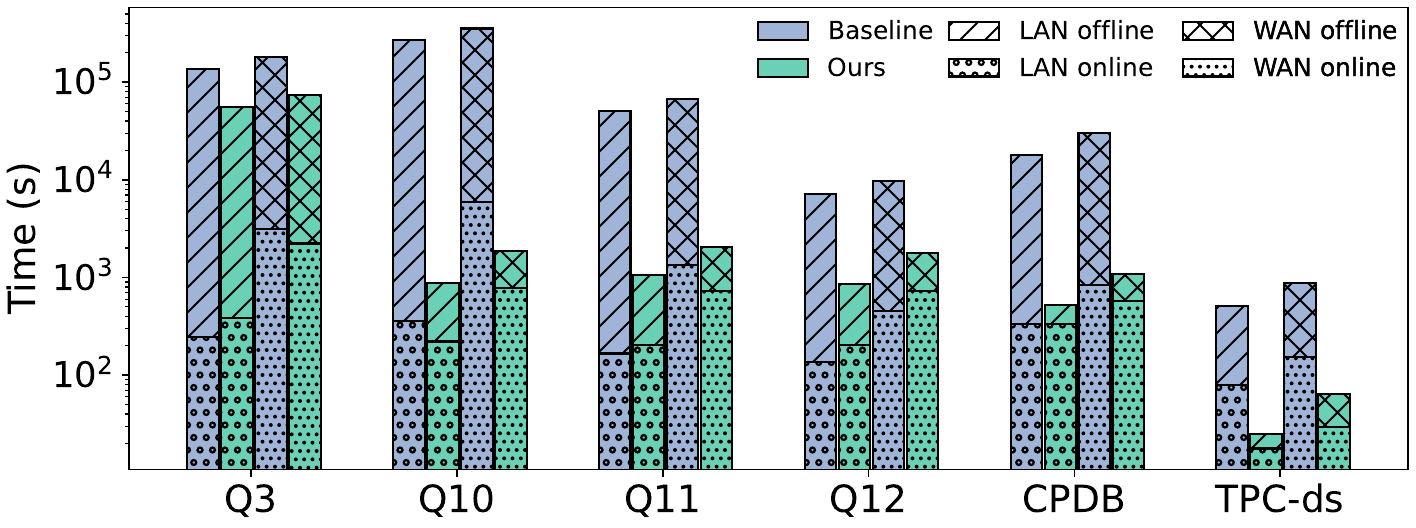}
    \vspace{-0.1in}
    \caption{Evaluation on real-world databases, where the data size for TPC-H queries is $n=2^{20}$.}
    \vspace{-0.15in}
    \label{fig:realQuery}
\end{figure}

\partitle{Simulation} We evaluate the overall execution time of MapComp in both LAN and WAN environments.
We treat the join process as a materialized join view generation process and select the CPSI + $\functionsym{trav}$ and $\functionsym{aPSI}$ + $\functionsym{trav}$ as the baseline and our approach, respectively.
For GA, we choose $\protocolsym{oSorting}$ for Q3 and the one-side version of $\protocolsym{oSorting}$ for Q10, Q11, Q12, CPDB, and TPC-ds as our approach. 
We compare the execution time of online and offline (\eg multiplication triples generation used in MPC) and present the result in Fig.~\ref{fig:realQuery}.



From the results of Q3 in the LAN setting, we observe that our approach gives {$2.46\times$} overall time speed up over the baseline. It is mainly because the shorter input of $\protocolsym{oSorting}$ requires fewer MPC operations. A similar observation can be drawn from the WAN setting.
For the other 3 queries from the TPC-H benchmark, the total execution of our protocol is $2.46\times\sim308.9\times$ faster than the baseline in the LAN setting, and $2.44\times \sim 191.0\times$ faster in the WAN setting. This is because our protocol is sorting-free in the one-side case due to the plaintext brought by our view.
Moreover, thanks to our payload-independent view and utilization of oblivious switching, for Q10, Q12, and queries on CPDB and TPC-ds, our solution requires no time-consuming general circuit evaluation (hence no need to generate multiplication triples for MPC in the offline phase) and obtains enhanced offline and online evaluation efficiency. 
The evaluation results demonstrate the great advantage of our GA protocol over the baseline in real-world queries, yielding up to $308.9\times$ efficiency improvement. To confirm the efficiency improvements of \oursys in multiple query settings, we provide additional experimental results in Appendix \ref{sec:app_multi_query}. 

\iffullversion


\fi

\vspace{-0.0in}
\section{Discussion}
\label{sec:discussion}
\vspace{-0.0in}

\vspace{-0in}


\vspace{-0.05in}
\partitle{Extend for malicious security}
The building blocks of \oursys (\eg random shuffling and CPSI) have corresponding malicious versions \cite{wang2017authenticated,frederiksen2013fast,song2023secret,yang2024fairsec}, which means that our protocol can be extended to defend against malicious adversaries. This extension would include additional security measures to mitigate the risks posed by malicious adversaries. The detailed formal constructions and rigorous proofs are left as future works.

\vspace{-0.05in}
\partitle{Update on join keys} 
Our view refresh focuses on the efficient update of data payloads. When join keys are updated, \oursys can rebuild the view to support this case as the previous non-view works. It is worth mentioning that this happens relatively rarely in practice, since the value of a key in a table is typically rarely changed to ensure the logical relations between tables and the high efficiency of indexing \cite{databaseSystemConcept}. Our additional observation is that when join keys are updated (under modification or addition), the mapping relation in the join index-based materialized view \cite{valduriez1987join,o1995multi} is broken. This means the lower bound of the complexity of view refreshing under join key updating could be $O(n)$. This result might provide some insights that can be utilized by future studies to enable richer support for designing secure materialized views.

\section{Related work}
\vspace{-0in}
\label{sec:related}

\vspace{-0.05in}
\partitle{MPC-based collaborative analytics systems} 
SMCQL \cite{bater2017smcql,bater2018shrinkwrap} is one of the earliest database systems that target privacy for collaborative queries under the MPC model using garbled circuit and ORAM \cite{gentry2014garbled}. 
Hu-Fu \cite{tong2022hu} employs MPC to allow parties to query spatial databases in a data federation. Senate \cite{poddar2021senate} proposed a decomposition protocol to decompose MPC computation into paralleled-executed smaller units and provided security against malicious parties. Scape \cite{han2022scape} and SECRECY \cite{liagouris2021secrecy} are built on three computing parties and provide dedicated algorithms for common SQL operations such as join and aggregation. Some works employs differential privacy techniques to provide additional protection for results or intermediate data \cite{bater2018shrinkwrap,WangBNM22,dwork2014algorithmic,wang2025differential}.

\vspace{-0.05in}
\partitle{Private set operation-based approaches} A rich body of work exists on custom protocols for private set operations, \eg PSI \cite{chase2020private,bienstock2023near}, PSU \cite{jia2022shuffle,zhang2023linear}, Private ID \cite{buddhavarapu2020private,garimella2021private}, and Circuit PSI \cite{pinkas2018efficient,pinkas2019efficient}.
There are also many works aiming at specific functionalities over the result of the set operation \cite{laur2013oblivious,mohassel2020fast,li2021prism}. For example, Secure Yannakakis \cite{wang2021secure} supports collaborative join-aggregate queries with strong assumptions, which limits its use cases. PJC \cite{lepoint2021private} focuses on join-aggregate queries for the case that the data sizes are unbalanced.


\partitle{Federated computation} 
Federated computation enables multiple parties to collaboratively compute a prescribed function over their distributed data without exposing sensitive information \cite{schwermer2024federated}. This paper focuses on federated analysis, a specific application of federated computation where the target functions are database queries \cite{zhang2024fedknn}. Beyond analytics, the principles of federated computation have been applied to other fields, including federated learning \cite{mcmahan2017communication,yang2019federated, miao2023task,miao2025federated}, federated search \cite{wang2024feb4rag,fan2025fedvse}, and federated recommendation \cite{yang2020federated,zhang2024preventing}.


\vspace{-0in}
\section{Conclusion}
\label{sec:conclusion}
\vspace{-0in}

This paper proposes MapComp, a novel view-based framework that facilitates secure JGA queries. By employing a specially crafted materialized view, MapComp pre-computes and reuses duplicate, heavy workloads of joins to enhance overall execution speed. The materialized view enjoys payload independence, enabling further optimization of GA protocols and outperforming existing methods in various scenarios. Our idea of introducing join-index-based materialized views into secure queries and integrating sorting and bitmap encoding may inspire further research into the application of views to enable real-time updating of MPC databases. 
Looking ahead, we propose two directions to inspire future work: 1) expand MapComp to accommodate more general scenarios involving multiple data owners, and 2) design a materialized view that supports MPC-free data updating for more relational operators. 
\balance
\bibliographystyle{IEEEtran}
\bibliography{sample}

\clearpage

\appendix

\iffullversion

\subsection{Frequently Used Notations.}
\label{app:notations}

We demonstrate frequently used notations in Tab. \ref{tab:notation}.

\begin{table}[t!]
    \centering  
    \renewcommand{\arraystretch}{1.1}
    \begin{tabular}{| c  p{6cm} |}
        \hline
        Symbol & Description \\
        \hline
        $n$ & usually denoted as the number of tuples in a table or the length of a vector \\
        \hline
        $l$ & usually denoted as the bit length of some values \\
        \hline
        $\perp$ & $\binaryValue{null}$ \\
        \hline
        $[a,b]$ & $\{a, a+1, \ldots, b\}$ for $a \leq b$ \\
        \hline
        $ind(\phi)$ & an indicator function which outputs 1 when the predicate $\phi$ is true or $0$ otherwise \\
        \hline
        $X$ & a vector with $|X|$ elements, and the length is also denoted as $n_x = |X|$ \\
        \hline
        $\pi$ & an injective function permutation \\
        \hline
        $Y = \pi \cdot X$ & apply $\pi$ on a vector $X$ and output $Y = (x_{\pi(1)}, \ldots, x_{\pi(n)})$ \\
        \hline
        \hline
        $\tablesym{R}$ & a table with cardinality $|\tablesym{R}|$ \\
        \hline
        $\tablesym{R}[v]$ & the vector containing all values of
the attribute $v$ in $\tablesym{R}$ \\
        \hline
        $\tablesym{R}_i$ & the $i^{th}$ tuple of $\tablesym{R}$ \\
        \hline
        $\tablesym{R}_i[v]$ & the value of attribute $v$ in the $i^{th}$ tuple of $\tablesym{R}$ \\
        \hline
        $\domain{g_0}$ & the domain space of an attribute $g_0$ \\
        \hline
        $d_0$ & the size of a domain $\domain{g_0}$ \\
        \hline
        $\domain{g_0}_i$ & the $i^{th}$ distinct value in $\domain{g_0}$, $1\leq i \leq d_0$ \\
        \hline
        $\aggfun{agg}(v)$ & a aggregate function preformed on an attribute $v$ including $\aggfun{max}$, $\aggfun{min}$, $\aggfun{count}$, and $\aggfun{sum}$ \\
        \hline
        $\tablesym{R^0} \Join_k \tablesym{R^1}$ & $\tablesym{R^0}$ inner equal join $\tablesym{R^1}$ on condition $\tablesym{R^0}.k = \tablesym{R^1}.k$ \\
        \hline
        $\mathcal{G}_{g, \aggfun{agg}(v)}\tablesym{R}$ & shorthand for ``\textbf{Select} $g, \aggfun{agg}(v)$ \textbf{From} $\tablesym{R}$ \textbf{Group by} g'' \\

        \hline
        \hline
        $\party_u$ & a computing party and $u\in\{0, 1\}$ \\
        \hline
        $\share{x}$ & the secret share of a value $x$ \\
        \hline
        $E$ & the intersection flag vector \\
        \hline
        $\mathcal{V}_u$ & the view held by the party $\party_u$ \\
        \hline
        $\tablesym{J}^u$ & the re-ordered data transcript of $\tablesym{R}^u$ \\
        \hline
        $\share{\tablesym{T}}$ & usually denoted as a secret shared table \\
        \hline
    \end{tabular}
\caption{Frequently used notations in this paper.}
\label{tab:notation}
\end{table}

\subsection{Oblivious permutation on a plain vector (\texorpdfstring{$\protocolsym{perm}^\mathsf{p}$}{Lg} and \texorpdfstring{$\protocolsym{invp}^\mathsf{p}$}{Lg})}
\label{appendix:plainPerm}

When applying a shared permutation (or inverse permutation) on a plain vector $X$, it can be realized with concretely smaller cost compared to $\protocolsym{perm}^\mathsf{s}$ (or $\protocolsym{invp}^\mathsf{s}$). 
Specifically, to permute a length-$n$ plain vector $X$ owned by $\sender$ with a length-$n$ shared permutation $\share{\pi}$, $\protocolsym{perm}^\mathsf{p}$ proceeds as follows:

\begin{enumerate}
    \item The parties invoke $\functionsym{osn}^\mathsf{s}$, where $\receiver$ acts as the receiver with a random permutation $\sigma$. $\share{\rho} \leftarrow \functionsym{osn}^\mathsf{s}(\sigma, \share{\pi})$.
    \item The parties reveal $\share{\rho}$ to $\sender$, and $\sender$ computes $X' = \rho \cdot X$. The parties invoke $\functionsym{osn}^\mathsf{p}$, where $\receiver$ acts as the receiver with $\sigma^{-1}$. $\share{Y} \leftarrow \functionsym{osn}^\mathsf{p}(\sigma^{-1}, X')$.
\end{enumerate}

Similarly, $\protocolsym{invp}^\mathsf{p}(\share{\pi}, X)$ proceeds as follows:

\begin{enumerate}
    \item The parties compute $\share{\rho} \leftarrow \functionsym{osn}^\mathsf{s}(\sigma, \share{\pi})$ where $\sender$ acts as the receiver with a random permutation $\sigma$. $\sender$ compute $X' = \sigma\cdot X$.
    \item The parties reveal $\share{\rho}$ to $\receiver$. The parties compute $\share{Y} \leftarrow \functionsym{osn}^\mathsf{s}(\rho^{-1}, X')$ where $\receiver$ acts as the receiver with input $\rho^{-1}$.
\end{enumerate}

The correctness and security hold, similar to the oblivious permutation on the shared vector.

\subsection{Optimizations of \texorpdfstring{$\protocolsym{sorting}$}{Lg}}
\label{appendix:optiSorting}

\subsubsection{Optimized Sorting-based GA Protocol 
\texorpdfstring{$\protocolsym{oSorting}$}{Lg}}
\label{appendix:oSorting}

\begin{figure}[t!]
    \framebox{\begin{minipage}{0.97\linewidth}
        \begin{trivlist} 
            \item \textbf{Input:} 
            Same as Fig.~\ref{protocol:sortbasedgrouppre}.	
            
            \item \textbf{Protocol After step 2 of Fig.~\ref{protocol:sortbasedgrouppre}:} 
            \begin{enumerate}[start=3]

                \item The parties invoke $\functionsym{osn}^\mathsf{s}$ and append results into $\tablesym{T^{(1)}}$, where $\bob$ acts as receiver with input $\sigma_b$. $(\shareb{\tablesym{T^{(1)}}[e]}, \share{\tablesym{T^{(1)}}[g_0]}) \leftarrow \functionsym{osn}^\mathsf{s}(\sigma_b, (\shareb{E}, \share{\tablesym{J^0}[g_0]}))$.

                \item Invoke stable sorting: $(\share{\pi_{g_0}}, (\shareb{\tablesym{T^{(2)}}[e]}, \share{\tablesym{T^{(2)}}[g_0]})) \leftarrow \functionsym{sSort}(\shareb{\tablesym{T^{(1)}}[e]}, \share{\tablesym{T^{(1)}}[g_0]})$.
            
                \item The parties invoke $\functionsym{perm}^\mathsf{p}$ with $\bob$ acts as sender, and append results into $\tablesym{T^{(2)}}$: $(\share{\tablesym{T^{(2)}}[g_1]}, \share{\tablesym{T^{(2)}}[v_1]}, \share{\rho})\leftarrow \functionsym{perm}^\mathsf{p}(\share{\pi_{g_0}}, (\tablesym{T^{(1)}}[g_1], \tablesym{T^{(1)}}[v_1], \sigma_b))$.
            
                \item The parties invoke $\functionsym{perm}^\mathsf{p}$ and append results into $\tablesym{T^{(2)}}$, where $\alice$ acts as sender: $\share{\tablesym{T^{(2)}}[v_0]} \leftarrow \functionsym{perm}^\mathsf{p}(\share{\rho}, \tablesym{J^0}[v_0])$.
                
            \end{enumerate}
            \item \textbf{Then:} Run the remainder after step 5 in Fig.~\ref{protocol:sortbasedgrouppre}.
        \end{trivlist}
    \end{minipage}}
    \vspace{-0in}
    \caption{Optimized sorting-based GA protocol $\protocolsym{oSorting}$.}
    \label{protocol:optSortbasedgrouppre}	
\end{figure}

We now demonstrate the protocol details of our final $\protocolsym{oSorting}$.
It is observed that performing shared permutation $\functionsym{perm}^s$ over a shared vector $\share{X}$ is expensive since it would invoke OSN for 4 times as mentioned in \S \ref{sec:obli_primitives}.
We note that $\functionsym{perm}^s$ can be avoided by invoking $\functionsym{perm}^\mathsf{p}$ (described in Appendix \ref{appendix:plainPerm}) that calls OSN only 2 times and has nearly half the communication cost compared with $\protocolsym{perm}^\mathsf{s}$. $\functionsym{perm}^\mathsf{p}$ applies shared permutation on a plain vector (instead of shared vector as $\protocolsym{perm}^\mathsf{s}$), so we optimize $\protocolsym{sorting}$ to obtain the final optimized sorting-based GA protocol $\protocolsym{oSorting}$ as showed in Fig.~\ref{protocol:optSortbasedgrouppre}. Correctness is guaranteed by the associative law of permutation.

\subsubsection{Optimization for \texorpdfstring{$\aggfun{sum / count}$}{Lg}}
\label{appendix:sumCount}
We observe that apart from $\functionsym{mux}$, the core operation in the underlying protocols of $\functionsym{trav}$ for $\aggfun{sum/count}$ is just local addition due to the property of additive secret sharing. Considering the aggregation result will be revealed directly, the aggregation can be further optimized to avoid prefix operation of $O(\log n)$ rounds. 
Taking $\aggfun{sum}$ as an example, the protocol can be modified as follows:
\begin{enumerate}
    \item Replace the computation in step 6: $\shareb{f_i} = \shareb{\tablesym{T^{(2)}_i}[e]} \odot \\(\lnot \functionsym{eq}(\share{\tablesym{T^{(2)}_i}[g_0] || \tablesym{T^{(2)}_i}[g_1]}, \share{\tablesym{T^{(2)}_{i-1}}[g_0] || \tablesym{T^{(2)}_{i-1}}[g_1]}))$.
    \item Replace step 7(a) with: $i\in[n]$:  $\share{\tablesym{T^{(3)}_i}[r_u]} = \sum_{j=i}^n{\share{\tablesym{T^{(2)}_j}[v_u]}}$.
\end{enumerate}
Moreover, $\bob$ needs to perform additional computation after revealing $\tablesym{T^{(4)}}$. $\bob$ picks out the tuples in $\tablesym{T^{(4)}}$ whose values of $g_0, g_1$ are not $\perp$, and sorts those tuples based on the lexicographical order of $g_0, g_1$ into $\tablesym{R}$. Then, $\bob$ updates $\tablesym{R_i}[r_u] = (\tablesym{R_i}[r_u] - \tablesym{R_{i+1}}[r_u])$ for $1 \leq i < |\tablesym{R}|, j\in\{0, 1\}$ as the result of $\aggfun{sum}$. In this way, the communication round of step 7 can be reduced into $O(1)$. This optimization can also be applied to $\protocolsym{bSorting}$ and $\protocolsym{mix}$ for group-$\aggfun{sum / count}$. 

\subsubsection{New Oblivious Stable Sorting Protocol $\protocolsym{bitSort}$ for Secret-shared Bitmap}
\label{appendix:bitSort}

\begin{figure}[t!]
    \framebox{\begin{minipage}{0.97\linewidth}
        \begin{trivlist} 
        
            \item \textbf{Input:} 
            Length-$n$ vectors $\shareb{B^1}, \dots, \shareb{B^d}$ with $\forall i\in[n]$: $\sum_{j = 1}^d{b^j_i}\in\{0, 1\}$.
            
            \item \textbf{Protocol:}
            \begin{enumerate}
                \item Initialize $\share{a} = 0$, length-$n$ vector $\share{V} = (\share{0}, \dots, \share{0})$.

                \item Compute $\shareb{B^{d + 1}}$: $\shareb{b^{d + 1}_i} = \lnot\bigoplus_{j = 1}^d{\shareb{b^j_i}} $ for $i\in[n]$.

                \item For $j = d + 1$ to $1$:
                \begin{itemize}
                    \item For $i = 1$ to $n$:
                    \begin{enumerate}
                        \item $\share{a} = \share{a} + \functionsym{b2A}(\shareb{b^j_i})$;
                        \item $\share{v_i} = \share{v_i} + \functionsym{mux}(\shareb{b^j_i}, \share{a}, \share{0})$.
                    \end{enumerate}
                \end{itemize}
                
            \end{enumerate}
            \item \textbf{Output:} $\share{V}$.
        \end{trivlist}
    \end{minipage}}
    \vspace{-0in}
    \caption{New oblivious stable sorting protocol $\protocolsym{bitSort}$ for secret-shared bitmap input.}
    \vspace{-0in}
    \label{protocol:bitmapSort}	
\end{figure}

We propose an efficient stable sorting protocol $\protocolsym{bitSort}$ and may be of independent interest. It inputs a secret-shared bitmap, which is $d$ length-$n$ binary shared vectors $\shareb{B^1}, \dots, \shareb{B^d}$ satisfying $\forall i\in[n]$: $\sum_{j = 1}^d{b^j_i}\in\{0, 1\}$, and outputs a length-$n$ shared permutation $\pi$. $\pi$ represents a stable sorting of $\shareb{B^1}, \dots, \shareb{B^d}$, such that the $i^{th}$ elements should be placed into $\pi(i)^{th}$ position in the sorted result. The protocol is shown in Fig.~\ref{protocol:bitmapSort}. It takes $O(1)$ rounds and $O(dn\log n)$ bits of communications, where $\log n$ is the bit length to represent the values in a length-$n$ permutation. Correctness and security follow from the Radix sort protocol in the three-party setting \cite{asharov2022efficient}.

\subsection{Configuration Details of the Simulated Queries}
\label{appendix:query}

\subsubsection{TPC-H}

The selected SQLs (Q3, Q10, Q11, and Q12) are 4 JGA queries from the wide-adopted TPC-H benchmark. They are designed to simulate real-world statistical queries and provide a fair comparison of efficiency. These queries are also adopted in many works \cite{han2022scape,liagouris2021secrecy,volgushev2019conclave,wang2021secure,poddar2021senate} to simulate complex real-world JGA SQL situations. 

We follow the paradigm \cite{bater2017smcql} that query execution be divided into local computation (the part of the query that involves only one party’s data) and the rest of the computation (involves both party's data and should be performed based on MPC). We illustrate it more concretely and taking Q3 as an example as follow.

\framebox{\begin{minipage}{0.9\linewidth}
\sqlword{\textbf{SELECT} $\mathsf{l\_orderkey}$, sum($\mathsf{award\_type}$ * (1-$\mathsf{l\_discount}$) as $\mathsf{revenue}$, $\mathsf{o\_orderdate}$, $\mathsf{o\_shippriority}$  \\
\textbf{FROM} 
$\tablesym{customer}$, $\tablesym{orders}$, $\tablesym{lineitem}$  \\
\textbf{WHERE} $\mathsf{c\_mktsegment}$ = '[SEGMENT]' and $\mathsf{c\_custkey = o\_custkey}$ and $\mathsf{l\_orderkey} = \mathsf{o\_orderkey}$ and $\mathsf{o\_orderdate}$ < date '[DATE]' and $\mathsf{l\_shipdate} >$  data '[DATE]' \\
\textbf{GROUP BY} 
$\mathsf{l\_orderkey,o\_orderdate, o\_shippriority}$ \\
\textbf{ORDER BY} 
$\mathsf{revenue}$ \textbf{DESC}, $\mathsf{o\_orderdate}$ 
}
\end{minipage}}

\vspace{0.1in}

The filter operations in the WHERE clause ($\mathsf{c\_mktsegment}$ = '[SEGMENT]' , $\mathsf{o\_orderdate}<$  date '[DATE]' and $\mathsf{l\_shipdate}>$ date '[DATE]' ) can be performed locally first. After this, the join condition ($\mathsf{c\_custkey}$ = $\mathsf{o\_custkey}$ and $\mathsf{l\_orderkey}$ = $\mathsf{o\_orderkey}$) and group-aggregation (sum and group-by) involve both parties' data, so they are performed by MPC-based MapComp to ensure end-to-end security. Specifically, the join keys ($\mathsf{c\_custkey}$ = $\mathsf{o\_custkey}$ and $\mathsf{l\_orderkey}$ = $\mathsf{o\_orderkey}$) in the dataset are input into view-generation protocol to create a materialized join view, then the sum (over ($\mathsf{l\_extendedprice*(1-l\_discount)}$) and group-by (over $\mathsf{l\_orderkey}$, and $\mathsf{o\_orderdate}$) are executed over the view with GA protocols. Q10, Q11 and Q12 are configured similarly as above and we omit the details.

\subsubsection{CPDB and TPC-ds}

Here, we present our simulated queries for CPDB and TPC-ds databases. The first query performes on CPDB to count the number of awards received by officers with complaints over a threshold NUM, grouped by the time and award type. The second query is to sum unsold products per warehouse and product category for a specified date DATE. 

\noindent \textbf{1. Query for CPDB:}

\framebox{\begin{minipage}{0.9\linewidth}
\sqlword{\textbf{SELECT} $\tablesym{A}$.$\mathsf{month}$, $\tablesym{A}$.$\mathsf{award\_type}$, count(*) \\
\textbf{FROM} 
$\tablesym{Award}$ A 
\textbf{JOIN} \\
(\textbf{SELECT} $\mathsf{UID}$ \textbf{FROM} $\tablesym{Complaint}$ \textbf{GROUP BY} $\mathsf{UID}$ \\
\textbf{HAVING} count($\mathsf{complainant\_id} >$) NUM) B
\\
\textbf{ON} $\tablesym{A}$.$\mathsf{UID}$ = $\tablesym{B}$.$\mathsf{UID}$ \\
\textbf{GROUP BY} $\tablesym{A}$.$\mathsf{month}$, $\tablesym{A}$.$\mathsf{award\_type}$
}
\end{minipage}}

\vspace{0in}
\noindent \textbf{2. Query for TPC-ds:}
\vspace{0in}

\framebox{\begin{minipage}{0.9\linewidth}
\sqlword{\textbf{SELECT} $\mathsf{i\_category}$, $\mathsf{warehouse}$, sum($\tablesym{A}$.$\mathsf{quanity}$) - sum($\tablesym{B}$.$\mathsf{quanity}$) \\
\textbf{FROM} \\
(\textbf{SELECT} $\mathsf{i\_id}$, $\mathsf{i\_category}$, $\mathsf{warehouse}$, $\mathsf{quanity}$ \textbf{FROM} $\tablesym{Inventory}$ \textbf{WHERE} $\mathsf{date}$ = DATE) A \\
\textbf{JOIN} \\
(\textbf{SELECT} $\mathsf{i\_id}$, $\mathsf{warehouse}$, sum($\mathsf{quantity}$) \textbf{FROM} $\tablesym{web\_sale}$ \textbf{WHERE} $\mathsf{date}$ = DATE \textbf{GROUP BY} $\mathsf{i\_id}$, $\mathsf{warehouse}$) B 
\\
\textbf{ON} $\tablesym{A}$.$\mathsf{i\_id}$ = $\tablesym{B}$.$\mathsf{i\_id}$ and $\tablesym{A}$.$\mathsf{warehouse}$ = $\tablesym{B}$.$\mathsf{warehouse}$\\
\textbf{GROUP BY} $\mathsf{i\_category}$, $\mathsf{warehouse}$
}
\end{minipage}}

\vspace{-0in}
\subsection{Additional experimental result}
\label{sec:app_multi_query}
\vspace{-0in}

\begin{table}[ht]
\vspace{-0in}
    \centering  
    \begin{tabular}{| c | c | c | c | c | c | c |}
        \hline
        \textbf{Num.} & \textbf{Q3} & \textbf{Q10} & \textbf{Q11} & \textbf{Q12} & \textbf{CPDB} & \textbf{TPC-ds}\\
	\hline
        \textbf{2} & 4.96 & 1170.47 & 154.97 & 31.23 & 129.46 & 73.68\\
        \hline
        \textbf{4} & 4.98 & 2122.87 & 227.3 & 58.51 & 239.01 & 128.15 \\
        \hline
        \textbf{6} & 4.99 & 2912.95 & 269.18 & 82.56 & 332.91 & 170.06 \\
        \hline
        \textbf{8} & 4.99 & 3578.95 & 296.5 & 103.92 & 414.3 & 203.31 \\
        \hline
    \end{tabular}
\caption{The ratio of the total execution time of baseline vs. \oursys when querying multiple times in LAN setting. Num. denotes the number of JGA queries.}
\vspace{-0in}
\label{tab:multipleQuery}
\end{table}

To confirm the performance gain of our \oursys dealing with multiple queries and payload dynamics, we conducted multiple JGA queries in the LAN setting, where the first query involves a view generation and subsequent queries require only an on-demand view refresh.
The setting of the queries, the baseline, and our approaches are the same as above. We quantified the ratio of the execution time between the baseline and MapComp in Tab.~\ref{tab:multipleQuery}. As the number of queries increases, the efficiency advantage of MapComp grows, achieving up to $3578.9\times$ efficiency improvement over the baseline. 
Similar trends can be observed in all queries. It can be primarily attributed to our more efficient GA protocols and view refreshing compared to the baseline. This result confirms the greater improvements of MapComp in cases of multiple queries under data dynamics, enabling the possibility to support a real-time updating database.


\else 
    \vspace{-0.05in}
\section{Brief Proof of Theorem~\ref{theorem:map}}
\label{proof:map}
\vspace{-0.1in}

\begin{proof} 
We show that there exists a PPT simulator $\simulator_0$ that can generate the simulated view given $\alice$’s inputs, which is statistically indistinguishable from the joint distribution of corrupted $\alice$’s view in the real execution of the protocol.
$\simulator_0$ can sample random strings $\shareb{\tilde{E}}_0, \share{\tilde{Z}}_0, \shareb{\tilde{F}'}_0, \shareb{\tilde{F}}_0, \share{\tilde{L}}_0, \share{\tilde{\sigma_0}}_0, \share{\tilde{\sigma_1}}_0, \share{\tilde{P}^0}_0, \\ \share{\tilde{P}^1}_0, \share{\tilde{\pi_1}}_0$ as the simulated view. 
It can be proven that the distributions of the simulated view and the real view of $\alice$ during the execution of $\protocolsym{aPSI}$ are indistinguishable in the hybrid world of the invoked functionalities.
Similarly, the simulator $\simulator_1$ for $\bob$ can be constructed, and the security of $\protocolsym{aPSI}$ is proven.
\end{proof}

\vspace{-0.25in}
\section{View for PK-FK join}
\label{sec:app_pk_fk_join}
\vspace{-0.1in}

Now, we describe our view design for the PK-FK join. The definition of the PK-FK view is slightly different from the PK-PK view, and the generation and refresh require additional steps. 

W.L.O.G, we assume $\bob$'s join key is a foreign key, which means the values of $\tablesym{R^1}[k]$ are non-unique, and we can divide them into groups based on distinct FK values. Our high-level idea is to first align a single tuple within an FK group with the corresponding tuple having the same PK key. Then, the payloads of PK tuples are obliviously duplicated to the correct locations to align the remaining tuples, completing the PK-FK join. The single-tuple alignment process is independent of the payload, which means it is reusable when the payload is updated, so the view refreshing is partially free. We illustrate the definition and operations of the PK-FK join view as follows.

\vspace{-0.2in}
\subsubsection{View for PK-FK join}

Given two tables $\tablesym{R}^0, \tablesym{R}^1$ with join key $k$, the views held by $\party_0, \party_1$ are 
$\mathcal{V}_0 =(\pi_0, \share{E}_0^b, \share{\tablesym{J}^0}_0)$ and $\mathcal{V}_1 =(\pi_1, \sigma, \share{E}_1^b, \share{\tablesym{J}^0}_1, \tablesym{J}^1)$:

\vspace{-0.1in}
\begin{enumerate}
    \item $\tablesym{J}^1 = \sigma \cdot \pi_1 \cdot \tablesym{R}^1$, and $e_i = 1$ iff $\tablesym{J}^1_i[k] \in \tablesym{R}^0[k]$.
    \item For $1 \leq i  < |\tablesym{J}^1|$, $\tablesym{J}^1_i[k] \leq \tablesym{J}^1_{i + 1}[k]$.
    \item For $i \in [|\tablesym{J}^1|]$: if $e_i = 1$, let $p = \mathsf{first}(\tablesym{J}^1[k], i)$, then $\tablesym{J}^0_p = \tablesym{R}^0_{\sigma \cdot \pi_0(p)}$ and $\tablesym{J}^0_i = \tablesym{J}^0_p$; if $e_i = 0$, $\tablesym{J}^0_i = \tablesym{R}^0_{\sigma \cdot \pi_0(i)}$.
\end{enumerate}

\vspace{-0.3in}
\subsubsection{Generation and Refresh}
\label{sec:pkfkGenRefresh}

The view generation proceeds as follows, and the view refresh only requires the last two steps, so it is partially free. Note that the cost of refresh is relevantly small since it only requires oblivious switching and oblivious traversal taking $O(1)$ rounds and $O(nl \log n)$ bits of communication.

\noindent \textbf{1. Mapping and alignment.}
First, we align a single tuple within an FK group with the corresponding tuple having the same PK key. To achieve this, a constant 1 is appended to the PK value by $\alice$ and a counter number $\mathbf{t}[s]$ (\eg 1,2,3...) is appended to the FK value by $\bob$ for each tuple $\mathbf{t}$, such that $\mathbf{t}[s]$ denotes an incremental counter of tuples with the same join key value $\mathbf{t}[k]$ (within the same FK group). 
Then, the parties invoke PK-PK view generation protocols (described in \S\ref{sec:viewGen}) with inputs $\{\mathbf{t}[k] || 1\}_{\mathbf{t}\in \tablesym{R^0}}$ and $\{\mathbf{t}[k] || \mathbf{t}[s]\}_{\mathbf{t}\in \tablesym{R^1}}$, respectively. $\alice$ obtain $\pi_0$, $\shareb{E}_0$ and $\bob$ obtain $\pi_1$, $\shareb{E}_1$. 
Finally, two parties reorder the databases with $\pi_0, \pi_1$ to obtain a temporary transcript $\tablesym{D}^i = \pi_i \cdot \tablesym{R}^i$.
The tuple $\mathbf{t}^1 \in \tablesym{D}^1$ with $\mathbf{t}^1[s] = 1$ will be aligned with a tuple $\mathbf{t}^0 \in \tablesym{D}^0$ with $\mathbf{t}^0[k] = \mathbf{t}^1[k]$ if $\mathbf{t}^1[k] \in \tablesym{D}^0[k]$; or a tuple $\mathbf{t}^0 \in \tablesym{D}^0$ with $\mathbf{t}^0[k] \notin \tablesym{D}^1[k]$ otherwise. At this point, the first tuple of each FK group of $\tablesym{D}^1$ is correctly joined with the corresponding PK tuple of $\tablesym{D}^0$.

\noindent \textbf{2. Local sorting and oblivious switch.}
$\bob$ sorts the table $\tablesym{D}^1$ based on the key attributes $k, s$ to get the permutation $\sigma$ and result table $\tablesym{J}^1$. The parties invoke $\functionsym{osn}^p$ to switch $\tablesym{D}^0, \shareb{E}$ with $\sigma$ and obtain $\share{\tablesym{J}^0}, \shareb{E'}$. After this step, the tuples of $\tablesym{J}^1$ with the same key will be mapped together and sorted by $s$. 

\noindent \textbf{3. Duplicate the tuples.} To achieve PK-FK alignment, the last step is to obliviously set the payload of remaining tuples of $\share{\tablesym{J}^0}$ as correct values. The parties obliviously duplicate the tuples of $\share{\tablesym{\tablesym{J}^0}}$, such that $\tablesym{J}^0_i = \tablesym{J}^0_{\mathsf{first}(\tablesym{J}^1[k], i)}$ holds if $e'_i = 1$, where $\mathsf{first} (\cdot, i)$ returns the first index of the group $i$.

\vspace{-0.07in}
\begin{enumerate}
    \item For $i \in |\tablesym{J}^0|$, $\share{\tablesym{J}^0_i} \leftarrow \functionsym{mux}(\shareb{e'_i}, \share{\tablesym{J}^0_i}, \share{\perp})$;
    \item $\shareb{E} \leftarrow \functionsym{trav}(\share{\tablesym{J}^1[k]}, \shareb{E'}, \aggfun{xor})$; $\share{\tablesym{J}^0} \leftarrow \functionsym{trav}(\share{\tablesym{J}^1[k]}, \share{\tablesym{J}^0}, \aggfun{sum})$.
\end{enumerate}
\vspace{-0.07in}

$\alice$ set $\mathcal{V}_0 = (\pi_0, \shareb{E'}_0, \share{\tablesym{J}^0}_0)$, $\bob$ set $\mathcal{V}_1 = (\pi_1, \shareb{E'}_1, \share{\tablesym{J}^0}_1, \tablesym{J}^1)$.
This is the desired PK-FK join view output, since for every valid tuple $\tablesym{J}^1_i$ that satisfies $e_i=1$, the payload of tuple $\tablesym{J}^0_i$ is correctly joined and aligned with it.

\begin{figure}[t!]
    \centering
    \includegraphics[width=0.8\linewidth]{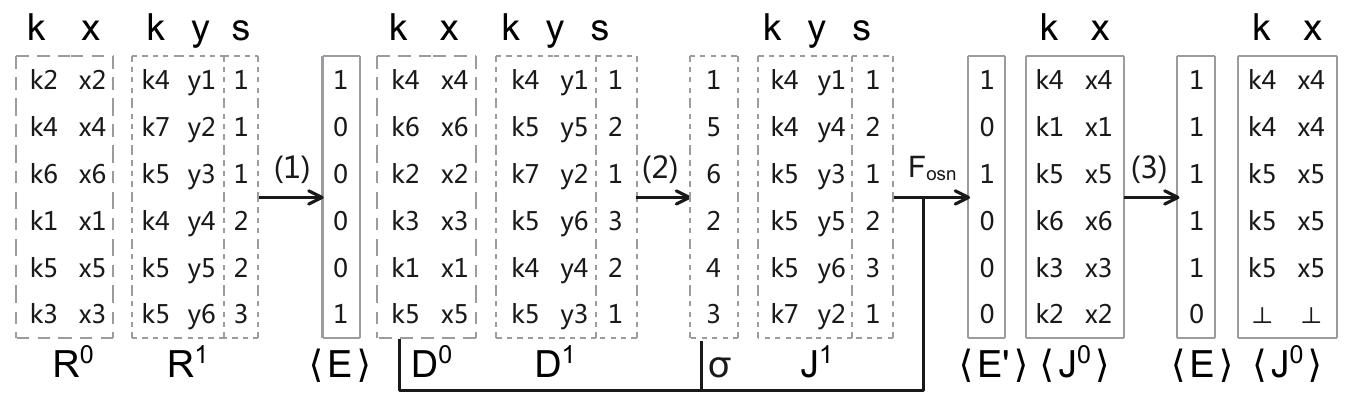}
    \vspace{-0.2in}
    \caption{Example of the view generation for the PK-FK join, where the result permutations in the first step are $\pi_0 = (2, 3, 1, 6, 4, 5)$ and $\pi_1 = (1, 5, 2, 6, 4, 3)$. We use $x, y$ to denote the other attributes besides $k$ in $\tablesym{R}^0$ and $\tablesym{R}^1$.}
    \label{fig:pkfk_view_exp}
    \vspace{-0.2in}
\end{figure}

\vspace{-0.2in}
\section{Additional experimental result}
\label{sec:app_multi_query}
\vspace{-0.1in}

To confirm the performance gain of our \oursys dealing with multiple queries and payload dynamics, we conducted multiple JGA queries in the LAN setting, where the first query involves a view generation and subsequent queries require only an on-demand view refresh.
The setting of the queries, the baseline, and our approaches are the same as above. We quantified the ratio of the execution time between the baseline and MapComp in Tab.~\ref{tab:multipleQuery}. As the number of queries increases, the efficiency advantage of MapComp grows, achieving up to $3578.9\times$ efficiency improvement over the baseline. 
Similar trends can be observed in all queries. It can be primarily attributed to our more efficient GA protocols and view refreshing compared to the baseline. This result confirms the greater improvements of MapComp in cases of multiple queries under data dynamics, enabling the possibility to support a real-time updating database. 

\begin{table}[ht]
\vspace{-0.2in}
    \centering  
    \begin{tabular}{| c | c | c | c | c | c | c |}
        \hline
        \textbf{Num.} & \textbf{Q3} & \textbf{Q10} & \textbf{Q11} & \textbf{Q12} & \textbf{CPDB} & \textbf{TPC-ds}\\
	\hline
        \textbf{2} & 4.96 & 1170.47 & 154.97 & 31.23 & 129.46 & 73.68\\
        \hline
        \textbf{4} & 4.98 & 2122.87 & 227.3 & 58.51 & 239.01 & 128.15 \\
        \hline
        \textbf{6} & 4.99 & 2912.95 & 269.18 & 82.56 & 332.91 & 170.06 \\
        \hline
        \textbf{8} & 4.99 & 3578.95 & 296.5 & 103.92 & 414.3 & 203.31 \\
        \hline
    \end{tabular}
\caption{The ratio of the total execution time of baseline vs. \oursys when querying multiple times in LAN setting. Num. denotes the number of JGA queries.}
\vspace{-0.25in}
\label{tab:multipleQuery}
\end{table}
\fi

\end{document}